\documentclass[useAMS, usenatbib, a4paper]{mn2e}

\usepackage{graphicx}
\usepackage{amssymb}
\usepackage[usenames,dvipsnames]{color}

\newcommand{\mrH}{\,\textit{r}_\mathrm{H}}

\newcommand{ \rHb}{\,\textit{r}$_\mathrm{h}$}
\newcommand{\mrHb}{\,\textit{r}_\mathrm{h}}

\newcommand{\mofbin}[1]{#1_\mathrm{B}}

\newcommand{\Ti}{\textit{T}$_{1}$}
\newcommand{\Tii}{\textit{T}$_{2}$}
\newcommand{\Tiii}{\textit{T}$_{3}$}

\newcommand{\tetai}{$\theta_{1}$}
\newcommand{\tetaii}{$\theta_{2}$}

\newcommand{\mtetai}{\theta_{1}}
\newcommand{\mtetaii}{\theta_{2}}

\newcommand{\refig} [1]{Figure~\ref{fig:#1}}
\newcommand{\refeq} [1]{Equation~\ref{eq:#1}}
\newcommand{\reftab}[1]{Table~\ref{tab:#1}}

\newcommand{\dgr}{\textordmasculine}

\newcommand{\corr}[2][ ]{#2}
\newcommand{\gram}[1]{#1}

\title[Capture configurations of binary-asteroids]{Irregular satellites of Jupiter: Capture configurations of binary-asteroids}
\author[H. S. Gaspar, O. C. Winter \& E. Vieira Neto]{H. S. Gaspar\thanks{E-mail: helton.unesp@gmail.com (HSG); ocwinter@feg.unesp.br (OCW); ernesto@feg.unesp.br (EVN)}, O. C. Winter\footnotemark[1] and E. Vieira Neto\footnotemark[1]\\ UNESP – Univ Estadual Paulista, Grupo de Din\^{a}mica Orbital e Planetologia, CEP 12.516-410,Guaratinguet\'{a}, SP - Brazil\\}

\begin{document}

\date{}

\pagerange{\pageref{firstpage}--\pageref{lastpage}} \pubyear{2009}

\maketitle

\label{firstpage}

\begin{abstract}
The origins of irregular satellites of the giant planets are an important piece of the giant ``puzzle'' that is the theory of Solar System formation. \corr[1]{It is well established that they are not \textit{in situ} formation objects, around the planet, as are believed to be the regular ones. Then, the most plausible hypothesis to explain their origins is that they formed elsewhere and were captured by the planet. However, captures under restricted three-body problem dynamics have temporary feature, which makes necessary the action of an auxiliary capture mechanism. Nevertheless, there not exist one well established capture mechanism}.%
\corr[NEW]{In this work, we tried to understand which aspects of a binary-asteroid capture mechanism could favour the permanent capture of one member of a binary asteroid.}%
We performed more than eight thousand numerical simulations of capture trajectories considering the four-body dynamical system Sun, Jupiter, Binary-asteroid. We restricted the problem to the circular planar prograde case, and time of integration to $10^4$ years. %
With respect to the binary features, we noted that 1) tighter binaries are much more susceptible to produce permanent captures than the large separation-ones. We also found that 2) the permanent capture probability of the minor member of the binary is much more expressive than the major body permanent capture probability. %
On the other hand, among the aspects of capture-disruption process, 4) a pseudo eastern-quadrature was noted to be a very likely capture angular configuration at the instant of binary disruptions. In addition, we also found that the 5) capture probability is higher for binary asteroids which disrupt in an inferior-conjunction with Jupiter. %
These results show that the Sun plays a very important role on the capture dynamic of binary asteroids.
\end{abstract}


\section{Introduction}
\gram{The existence of more than 350 natural satellites is known}, from which approximately 50\,\% are planetary ones. \corr[34]{An interesting point about this number, is that before 1997 just a tenth of such objects was known, i.e., the new ``CCD observational era'' allowed this number to increase by an order of magnitude within just a half decade \citep{gladmanetal98, gladmanetal00, gladmanetal01, sheppardjewitt03, holmanetal04, kavelaarsetal04, sheppardetal05, sheppardetal06}}. The planetary satellites can be distinguished into two characteristic groups: regulars and irregulars \citep{kuiper56, peale99}. The first group, is characterized by small values of semi-major axis, eccentricities and inclinations. These characteristics are a strong signature of \textit{in situ}-formation through matter accretion from the circumplanetary disc \citep{ luninestevenson82, vieiranetowinter01, canupward02, canupward06, mosqueiraestrada03, sheppardjewitt03}. In \gram{contrast}, the irregular satellites have large values of semi-major axis \citep{burns86}, \gram{often} high eccentricities and inclinations. A large part of irregular satellites have \gram{retrograde} orbital inclinations higher than 90 degrees \citep{jewitthaghighipour07}. Another important characteristic of the irregular ones are the family groups, i.e., satellite groups characterized by similar orbital elements \corr[5]{\citep{gladmanetal01, kavelaarsetal04}}. %
These peculiar characteristics are incompatible with the \textit{in situ}-formation model through matter accretion \citep{kuiper56}, and since they are the majority group of planetary satellites in the solar system, there exists a large scientific interest about their origin. Then, the most plausible hypothesis to explain their origins is that they formed elsewhere and were captured by the planet \corr[6]{\citep{kuiper56, heppenheimerporco77, pollacketal79, colombofranklin71}}. However, \gram{many studies} have shown that gravitational captures under three-body-dynamics are temporary \citep{everhart73, heppenheimerporco77, carusivalsecchi79, bennermckinnon95, vieiranetowinter01, wintervieiraneto01}. This fact has induced researchers to propose some auxiliary capture mechanism. Among others, we point out four mostly well known:
\begin{enumerate}
 \item Gas drag capture \citep{pollacketal79, cukburns04}: A \gram{temporarily} captured asteroid becomes permanently captured through kinetic energy decrease due to gas drag inside the circumplanetary disk of gas and dust;
 \item Pull-Down capture \citep{heppenheimerporco77, vieiranetoetal04, oliveiraetal07}: A temporary captured asteroid becomes permanently captured due to an increase of the Hill's radius of the planet. This increase in Hill's radius occurs due to either planet's mass growth or planet's migration away from the Sun;
 \item \gram{Close-approach} interaction captures \citep{colombofranklin71, tsui99, tsui00, astakhovetal03, nesvornyetal03, funatoetal04}: A \gram{temporarily} captured asteroid becomes permanently captured through energy and angular momentum exchanges with an existing satellite;
  \item Capture of binary-asteroids \citep{agnorhamilton06, vokrouhlickyetal08}: One member of a binary-asteroid becomes permanently captured when the binary approaches the planet and disrupts. 
\end{enumerate}

The capture mechanism of binary-asteroids is very interesting since the present observations have shown an increasing number of such systems in the main populations of such objects as the Kuiper Belt, Main Belt and Near Earth Asteroids \citep{noll06}. \citet{agnorhamilton06} presented numerical simulations of close encounters between Neptune and a binary-asteroid, where they considered \gram{one asteroid comparable} to Triton and \corr{a secondary}, with equal mass or one order of magnitude \corr{lower}. Their results show that is possible to \gram{disrupt} the binary when the close approach happens inside a spherical region whose radius, called tidal radius, is given by:
\begin{equation}
r_{td} = a_{B} \left( \frac {3 M_{P}}{m_{1}+m_{2}} \right)^{1/3}
\label{eq:eqrtd} 
\end{equation}
where $a_{B}$, $M_{P}$, $m_{1}$ and $m_{2}$ are the binary semi-major axis, planet mass, primary and secondary asteroid masses, respectively. A possible outcome after disruption is the capture of one member of the primordial binary-asteroid. 

The increasing number of binary-asteroid discoveries \citep{noll06} \gram{along with} the \citet{agnorhamilton06} results, \gram{have} motivated us to study the binary-asteroid capture process \gram{in the context of} four-body dynamic\gram{s}, where we considered Sun, Jupiter and a pair of asteroids. The purpose of this work is to identify the most important orbital characteristics inherent to the binary-asteroid capture/disruption process \gram{that produce} the permanent capture of at least one member. \gram{Compared to existing works }\citep{agnorhamilton06, vokrouhlickyetal08}, \gram{our study considers} the inclusion of solar perturbation. We found that the Sun's presence has a crucial influence on the binary capture/disruption process, at least in \gram{the} planar case.

This paper is built with the following structure: Section \ref{sec:capturemodeling} describes the model we use in our study as well the adopted numerical approach. Section \ref{sec:results} presents the results with an analysis of them. Finally, the last section summarizes our conclusions.

\section{Capture model}
\label{sec:capturemodeling}
Given the temporary feature of captures \gram{in the} three-body problem, we have chosen to perform a study under four-body dynamics using the Sun and Jupiter as primary bodies and a binary-asteroid. We basically propose a capture model in which a binary-asteroid first becomes temporarily captured by Jupiter, \gram{and then} disrupts and has one of its member permanently captured by Jupiter while the other one escapes. \corr[NEW]{The present paper addresses the early results of a more general work which is under development. It should be noted that the main goal of the present work is not to reproduce the actual configuration of Jupiter's irregular satellites, but rather, to comprehend how specific configurations can lead an asteroid member of a primordial binary to a permanent capture.}
As a first \gram{stage, we have} considered only the planar prograde case. Given this capture model, our task is to search for the initial conditions which yields asteroid temporary captures by Jupiter. The temporary captures, as well as the close encounters, are intrinsic features of solar system formation  \corr[6]{theories \citep{pollacketal96, hahnmalhotra05, tsiganisetal05, gomesetal05}}. Furthermore, temporary captures seem to be a more efficient mechanism to accomplish asteroid captures \gram{because there is a} longer interaction time between the binary-asteroid and the planet rather than a single passage with a very short time of interaction.
\subsection{Procedure}
\corr[8]{In this work, as a first study, we considered only the coplanar four-body dynamics. Furthermore, we set Jupiter's eccentricity to zero for all the simulations.} In order to perform the numeric \gram{studies} we used an integrator based on Gauss-Radau spacing \citep{everhart85}. In order to verify the integrator's accuracy we checked whether the system's total energy holds throughout the integration. We found the energy variation was lower than $10^{-11}$. \corr[9]{In addition, we have checked the value of the Jacobi constant of the binary-asteroid center of mass before binary disruptions for all the trajectories. We found that the Jacobi constant variation was lower than $10^{-9}$}

The adopted procedure basically followed three phases:
i) Firstly, we performed a capture time analysis of the system Sun-Jupiter-particle through which we obtain the, from now on designate, \textit{suitable initial conditions}, i.e., initial conditions which result in a particle's temporary capture by Jupiter. ii) Given the suitable initial conditions, we replace the individual particle by a \gram{pair} of bodies in order to set up a binary-asteroid, which will be called \textit{initial conditions}. iii) Finally, we study the binary-asteroid capture through simulations of the system Sun, Jupiter, binary-asteroid, by considering a set of different initial conditions derived from each one of the suitable initial conditions. 

\subsection{Suitable Initial Conditions}
\label{subsec:Capturetime}
In order to obtain the suitable initial conditions, we performed a primary capture time study following the steps of \citet{vieiranetowinter01}. It consists of the integration of a particle trajectory using a negative time step under the three-body dynamics considering Sun and Jupiter as primaries. By setting the particle to start orbiting around Jupiter we have three possible outcomes depending on particle's initial condition: i) The particle collides with Jupiter. ii) The particle remains orbiting the vicinity of Jupiter up to the final time of integration ($10^4$ years), in such cases, the particle's initial conditions are stable ones. iii) The particle escapes from Jupiter's vicinity and begins to orbit the Sun. 

\begin{figure}
	\includegraphics[width=0.48\textwidth, viewport=14 0 390 208, clip]{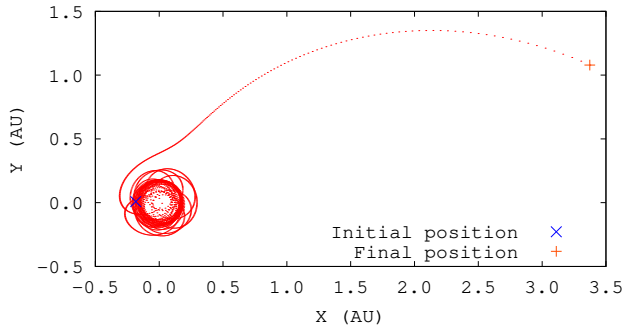}
	\caption{Example of escape trajectory in backward integration. \corr{The particle is in clockwise trajectory under Jupiter fixed frame of reference}
 The initial point indicated by blue \textquoteleft{x}\textquoteright\ corresponds to the primary initial conditions (around Jupiter). The final point indicated by red \textquoteleft{+}\textquoteright\ corresponds to the suitable initial condition, which will be replaced by the binary-asteroid.}
	\label{fig:Escape}
\end{figure}

In order to make easy the comprehension, we finally define these particle's orbital elements around Jupiter as \textit{primary initial conditions}. Summarizing:
\begin{description}
 \item Primary initial conditions: Three-body problem initial conditions which are backward integrated in time in order to obtain the \textit{suitable initial conditions};
 \item Suitable initial conditions: Three-body problem initial conditions which lead the particle to temporary captures by Jupiter. By replacing the particle for a binary-asteroid one derives the \textit{initial conditions} of the binary-asteroids;
 \item Initial conditions: The real initial conditions used to study the binary-asteroid capture dynamics, in which we consider Sun, Jupiter and a pair of asteroids;
\end{description}
We are particularly interested on the data relative to escape trajectories, which are capture ones when it is considered time forward. The particle's and Jupiter's orbital elements with respect to the Sun at the instant when the particle escapes\footnote{\corr[10]{We check the particle's two-body energy every time interval of one year during the numerical simulations. We consider that the escape} occurs when the particle's two-body energy with respect to Jupiter becomes positive.} are taken as the suitable initial conditions, as illustrates the example in \refig{Escape}.

\begin{figure}
\begin{flushright}
	\includegraphics[width=0.46\textwidth,viewport=29 35 371 355, clip]{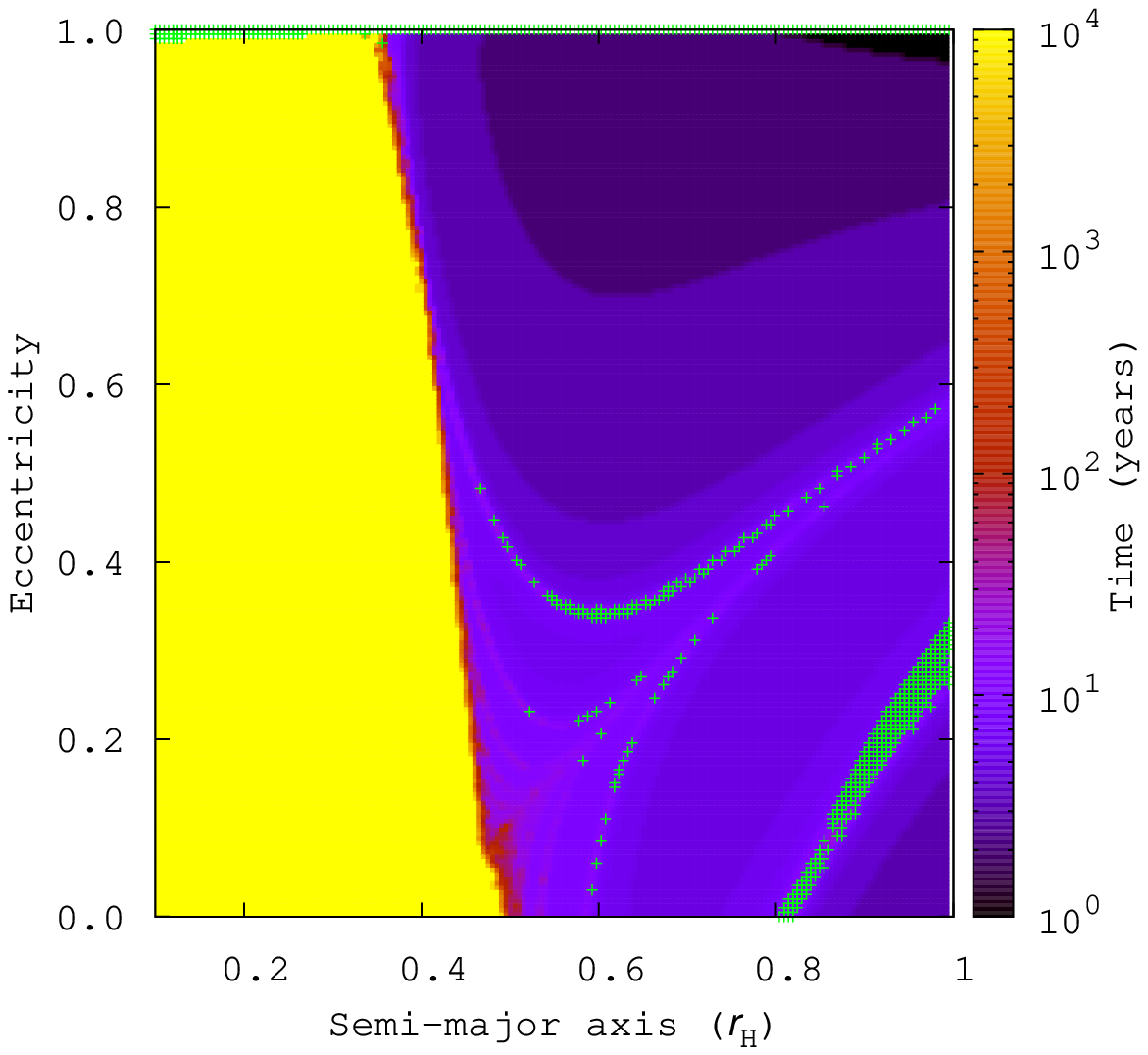}\\
	(a)\\
	\includegraphics[width=0.46\textwidth,viewport= 0 10 339 300, clip]{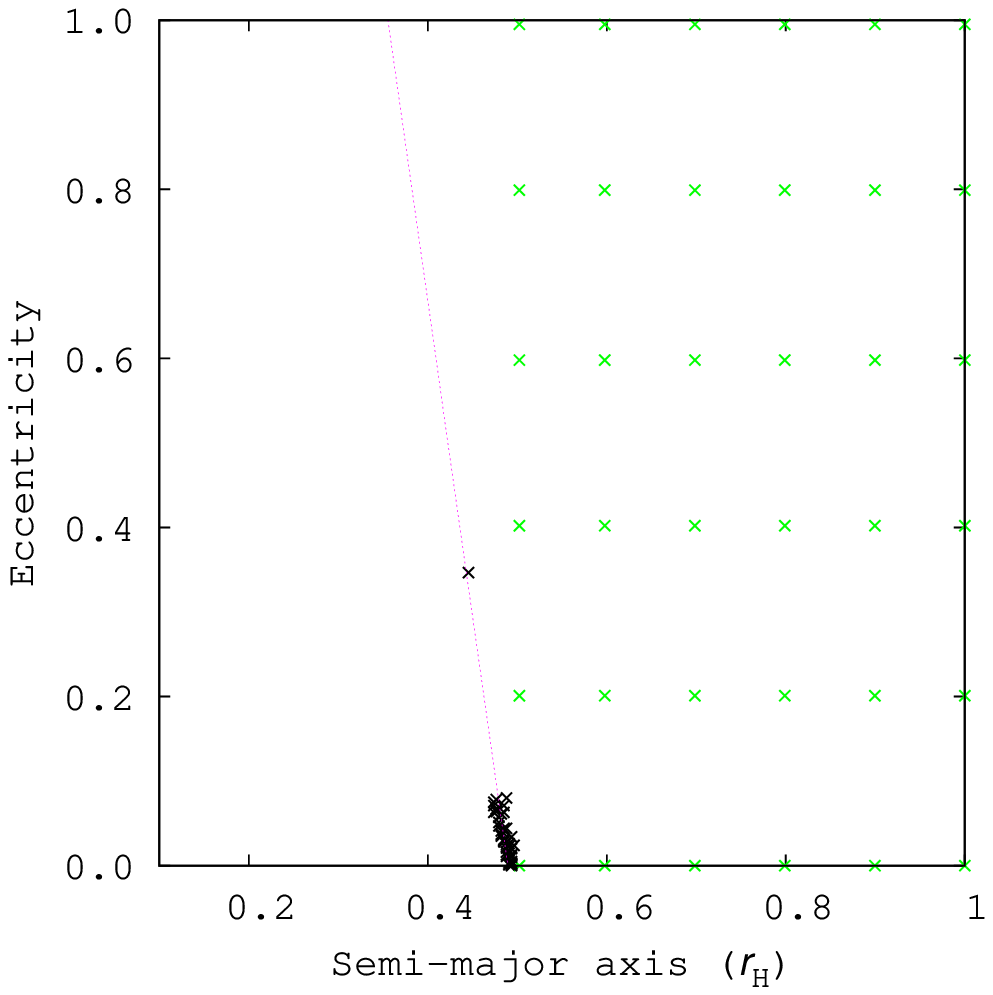}\\
	(b)
\end{flushright}
	\caption{(a) Capture time mapping. The color scale correspond to the capture time (in years) for such pair of initial semi-major axis and eccentricity. The green \textquoteleft+\textquoteright\ indicates the initial conditions which produce collisions.
        Initials longitude of pericenter and true anomaly of the particle were set to zero ($\varpi=f=0$). (b) Primary initial conditions used to obtain the suitable initial conditions. The green and black \textquoteleft{x}\textquoteright\ indicate the cases whose capture time are shorter and longer than a thousand of years, respectively. The pink dotted line is given by \refeq{limiteRita}.}
	\label{fig:Mapa}
\end{figure}

The main results of our three-body dynamics simulations, Sun-Jupiter-particle, is \gram{shown in the capture time map of \refig{Mapa}(a)}. \gram{This} is an $ a{\times}e$ diagram whose color scale \gram{denotes} particle's \gram{escape} time (in years) in the backward's integration. In this plot, the particle's semi-major axis \gram{is given} in terms of Hill's radius of Jupiter. This map has been generated by setting the particle's initial longitude of pericenter $(\varpi)$ and initial true anomaly $(f)$ \gram{with respect to Jupiter} as zero.

In plot (a) of \refig{Mapa}, the yellow region corresponds to the primary initial conditions which do not result in escapes from Jupiter in $10^4$ years. \corr[NOVO]{Considering the satellite eccentricity up to 0.5, \citet{domingosetal06} obtained} an expression \gram{for} this stable region as a critical semi-major axis inside which the satellites would remain stable. This expression is given by
\begin{equation}
 a_\mathrm{E} \approx 0.4895 (1.0000-1.0305e_\mathrm{P}-0.2738e_\mathrm{sat})
 \label{eq:limiteRita}
\end{equation}
where $e_{\mathrm{P}}$ and $e_\mathrm{sat}$ are the planet's and satellite's eccentricities, respectively. Particularly, in this work, \gram{the second term between parenthesis on the right hand side of \refeq{limiteRita} vanishes since} we have taken $e_{\mathrm{P}}=0$. In order to obtain this expression \citet{domingosetal06} followed the same procedure used by \citet{vieiranetowinter01} and also set the particle's initial longitude of pericenter $(\varpi)$ and initial true anomaly $(f)$ as zero. \corr{ In section \ref{subsec:ResultsTiii} we will show that the stable region can be more extensive for initial values of $\varpi$ and $f$ different from zero.}

\begin{figure}
 \includegraphics[bb=55 60 460 313, width=0.48\textwidth]{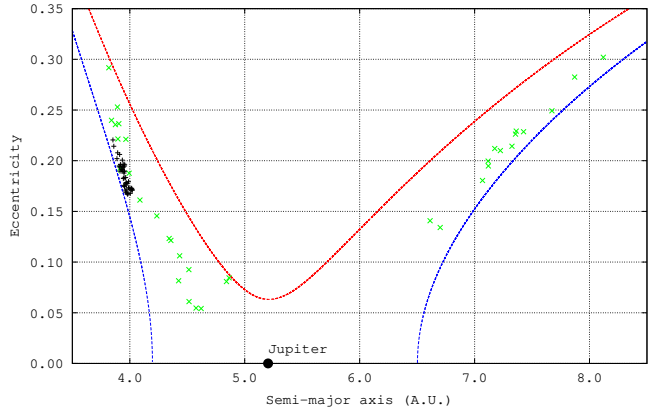}
 \caption{\corr[11]{Suitable initial conditions} $a \times e$ heliocentric diagram. The green ``x'' and black ``+'' indicate suitable initial conditions which were obtained from primary initial conditions whose capture time are shorter and longer than a thousand years, respectively. Red and blue curves are Tisserand relations $T=2.996$ and $T=3.036$, respectively, which encompass all the suitable initial conditions.}
 \label{fig:SuitableIC}
\end{figure}

Finally, the non-yellow region on the map (a) of \refig{Mapa} indicates the primary initial conditions \gram{that} resulted in escape. These are the cases from which we can take the suitable initial conditions. However, it is not feasible to use all the data obtained from the capture time analysis. By analyzing the capture times of escape cases, we found \gram{only 45 cases} in which the capture time exceed one thousand years. It \gram{is} feasible to take all these long time capture cases to compose the set of suitable initial conditions. Among the cases in which the capture times are shorter than a thousand of years, we took an \gram{uniformly-spaced} grid of points in $a\times e$ space in order to complete the set of suitable initial conditions with a representative set of the whole data. Plot (b) in \refig{Mapa} summarizes the set of primary initial conditions we used to obtain the suitable initial conditions. 
 The $a \times e$ diagram of \refig{SuitableIC} shows the suitable initial conditions which was obtained from the primary ones. The plot shows the semi-major axis and eccentricities of the main asteroid in the heliocentric frame. Furthermore, all the trajectories are direct with respect to the Sun. It is also plotted, in \refig{SuitableIC}, the \emph{Tisserand relations} for $T=2.996$ and $T=3.036$. The Tisserand relation \citep{tisserand1896}, as one can find in \citet{murray99}, is given by:
\begin{equation}
 T=\frac{1}{2a} + \sqrt{a(1-e^2)} \cos(I) \approx \mathrm{constant}
\end{equation}
where $a$, $e$ and $I$ are the object heliocentric semi-major axis, eccentricity and Inclination.


\subsection{Model for the Binary Asteroid}

The term \ binary-asteroid\\ \gram{can} refer to either a system in which a \gram{pair} of asteroids of similar masses orbit their common barycenter or an asteroid that has a small satellite \citep{noll06}. In order to compose a binary-asteroid by using the suitable initial conditions, we added a second body orbiting the first one. From now on, the main and the secondary asteroids will be called P1 and P2. For simplicity, we model a binary-asteroid by setting P2's orbital elements with respect to P1. In other words, asteroid P2 initially orbits asteroid P1. P2's orbital elements with respect to P1 will be referred as \textit{binary's elements}. Given our particular interest in Jupiter's irregular satellites, we set the P2's mass as $m_2=10^{19}\,kg$ based on Himalia's mass, the largest irregular satellite of Jupiter \citep{emelyanov05}. Using the same mass ratio as \citet{agnorhamilton06}, we set P1's mass $m_1=10\,m_2=10^{20}\,kg$. 

\corr[NEW]{As already stated, the main goal of this paper is to identify the most appropriate configurations that would generate permanent capture of one asteroid from a binary system.}  \corr[12a, 13a]{Thus, since we have a huge range of possibilities, is this paper we made some restrictions in order to be able to explore a significant part of the initial conditions space. Among the restrictions, we considered that P2 is always initially in prograde circular orbit around P1, $(e=I=0)$.}

From each of the 81 suitable initial conditions we derived 108 new initial conditions. We vary the initial binary's true anomaly $\mofbin{f}$ from 0 up to 330\textordmasculine\ \gram{in steps of} 30\textordmasculine, and the initial binary's semi-major axis $\mofbin{a}$ from 0.1\rHb\ up to 0.5\rHb\ \gram{in steps of} 0.05\rHb. Here \rHb, \corr[12b]{with `h' written in lowercase}, is P1's Hill's radius with respect to the Sun calculated for each one of the suitable initial conditions. \corr[12c]{Let's make clear that we did not make use of the tidal radius given by \refeq{eqrtd}}.The Hill's radius, as one can find in \citet{murray99}, is defined by:
\begin{equation}
r_{Hill} = \left( \frac{\mu}{3} \right)^{1/3} a,
\label{eq:hillradius}
\end{equation}
where $\mu$ and $a$ are the mass ratio and the semi-major axis, respectively.
\corr[13b]{The chosen upper semi-major axis limit of 0.5 \rHb\ is a well-established limit of stability for prograde systems \citep{hamiltonburns91, hamiltonburns92, domingosetal06}. The lower semi-major axis limit was arbitrarily chosen, though tighter binaries do exist.}
\corr[15]{By calculating the Hill's radius of P1 \rHb\ with respect to the Sun (\refeq{hillradius}), one finds that these values vary from $1 \times 10^{-3} AU$ to $2 \times 10^{-3} AU.$, and P2's initial orbital velocity with respect to P1 vary from $7 m/s$ to $22 m/s$.}

\subsection{Binary asteroid's capture simulations}

Since we derived 108 initial conditions from each one of 81 suitable initial conditions, we performed a total of 8\,748 binary-asteroid capture trajectory simulations. We set the trajectory integration time to 10$^4$\,years and the output time step to 10\,hours. In order to avoid a large amount of data storage we designed an algorithm which identified the integration main stages. We labeled each instant of these main stages, as follow:

\begin{description}
	\item \Ti: instant when the binary-asteroid is first captured by Jupiter;
	\item \Tii: instant when the binary-asteroid disrupts;
	\item \Tiii: instant when only one member of the disrupted binary-asteroid escapes from Jupiter;
\end{description}

Once the algorithm identifies each one of the three instants, it stores the instantaneous system configuration data, as well as the integration instant, in three distinct files. By taking the difference between \Ti\ and either \Tiii\ or the final time of integration, we can compute the capture time for each case. This identification algorithm basically consists on a two-body energy check-up, every integration step, described as follow:

\gram{The }binary-asteroid approaches Jupiter in a quasi-\corr[16a]{parabolic} trajectory, given that it initially orbits the Sun. \corr[16b]{For all the cases considered in this study at least one asteroid two-body energy with respect to Jupiter was positive}. At the instant \corr{we found that} both P1's and P2's two-body energies with respect to Jupiter became negative the instant \Ti\ is identified. \corr[16c]{Although, those energies do not become negative at the same instant.}

Similarly, P2's two-body energy with respect to P1 is initially negative given that P2 initially orbits P1. \corr[17]{Thus, if P2's two-body energy with respect to P1 becomes positive, instant \Tii\ is identified. Otherwise we do not identify neither instant \Tii\ nor \Tiii. Rather it could happen a double capture, a mutual collision or a double escape. However, in all the cases considered in this study the binary disrupted or collide with each other.}

Therefore, after the binary-asteroid's capture by Jupiter, both P1's and P2's individual two-body energies with respect to Jupiter are negative. Furthermore, after the binary-asteroid's disruption, \gram{interactions between P1 and P2 become negligible}, allowing either P1 or P2 to individually escape from Jupiter. Finally, at the instant in which either P1's or P2's two-body energy with respect to Jupiter became positive
 the instant \Tiii\ is identified.

Three possible outcomes succeed \Tiii:

\begin{enumerate}
 \item The remaining asteroid collides with Jupiter, which characterizes a \textit{collision};
 \item The remaining asteroid escapes from Jupiter, which characterizes a \textit{double escape};
 \item The remaining asteroid \gram{stays} bound throughout the \gram{rest of the} integration time, which we characterize as a \textit{permanent capture};
\end{enumerate}
\section{Results}
\label{sec:results}
We show in this section some plots built with the data stored at instants \Ti, \Tii\ and \Tiii\ in three respective subsections. We will discuss some statistical results in the fourth subsection, and show some examples of capture trajectories in the last subsection.
\subsection{Analysis at the instant of binary capture (\Ti)}
\begin{figure}
	\includegraphics[width=0.48\textwidth,clip]{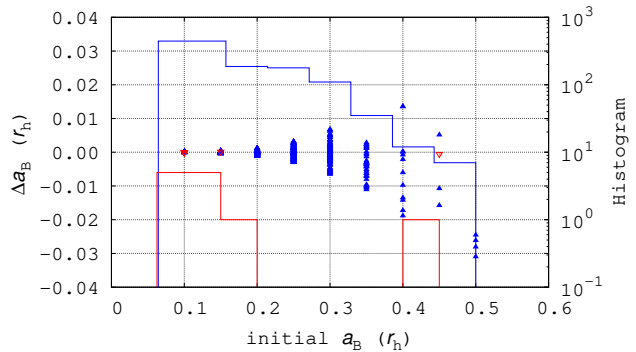}
	\caption{Binary's initial semi-major axis versus variation of the binary's semi-major axis at instant \Ti\ for the cases which resulted in permanent captures of either P1 or P2, in red and blue triangles, respectively. Red and blue filled lines are P1's and P2's permanent capture histograms, respectively.}
	\label{fig:TiCapaxDa}
\end{figure}

The plot in \refig{TiCapaxDa} compares the binary-asteroid separation at instant \Ti\ with its initial separation, through binary's semi-major axis $\mofbin{a}$ at instant \Ti\ and initial binary's semi-major axis $\mofbin{a}$, respectively. Both axis in this plot are measured in units of initial Hill's radius of P1 \rHb. This plot allows us to comprehend the binary-asteroid's evolution from the beginning of the integration to instant \Ti.

The histograms reveal that i) tighter binary-asteroids are more susceptible to permanent captures than binary-asteroids with large separations \gram{and} that ii) \corr[sub?]{the capture probability of the binary's smaller member is much higher than the major companion's capture probability.} 
\corr[NOVO]{ The blue histogram shows a roughly negative exponential behavior of permanent capture probability, with respect to the initial binary separation, which becomes very low for initial $\mofbin{a}\gtrapprox0.35\mrHb$}. Furthermore, beyond initial $\mofbin{a}=0.35\mrHb$ one can observe that the permanent capture occurs more often with binaries whose semi-major axis decreased. Consequently, \corr[NOVO]{these results point out to a limit $\mofbin{a}\approx0.4\mrHb$ beyond which permanent capture plausibility is negligible.} \corr[18a]{Evidently, $\Delta \mofbin{a}$ dispersion increases as the initial separations increases since it causes the binary bound to be weaker, consequently more susceptible to secular variations due to Sun and Jupiter perturbations.}

\corr[18a]{As weaker bounded binaries disrupt more easily, one could expect that they would more easily generate permanent captures. However, our results show that tighter binaries have higher probability to generate permanent captures. This apparent paradox can be understood in terms of the energy exchange needed to turn a temporary capture into a permanent one.} Based on \gram{results by} \citet{tsui99, tsui00}, which show that it is possible \gram{for} an asteroid to be kept captured by a planet due to exchange reactions with a local satellite, it is possible to explain the binary capture mechanism through energy exchanges in four steps:
\begin{enumerate}
 \item Lets consider a binary-asteroid, initially orbiting the Sun, which will be captured by Jupiter. Since the binary-asteroid is primordially orbiting the Sun, each one of its members' individual energy is higher than the escape energy $\varepsilon_0$, i.e., the minimum necessary energy to allow each asteroid individually escape from Jupiter.

 \item However, once the two asteroids orbits closely their common barycenter, angular momentum and energy exchanges occurs constantly. Furthermore, after the binary-asteroid be temporarily captured (\Ti), Jupiter starts to disturb the binary binding. Therefore, the exchange reactions become more intense;

 \item As a consequence of Jupiter's perturbation the binary-asteroid disrupts (\Tii). However, before the binary disruption the energy exchanges between the asteroids provides some energy states in which one member's energy is lower than the escape energy $\varepsilon_0$, which will not allow it to escape from Jupiter;

 \item Finally, after the binary disruption the interactions between the asteroids become negligible, so that, the asteroid whose energy decreased \corr{to values lower than $\varepsilon_0$} remains captured by Jupiter while the other whose energy increased will escape from Jupiter after some time (\Tiii).
\end{enumerate}

Note that \gram{these ideas} agree well with our results:
\begin{enumerate}
 \item \corr[18b]{The rupture of tighter binaries imply on larger energy exchange, implying on higher probability of permanent capture;}
 \item \corr[18b]{The smaller body of the binary is the one that suffers larger energy exchange and consequently is the one with higher probability of permanent capture;}
 \item Finally, the \corr{the existence of a} separation limit agrees well with the conclusions since the binary separation is proportional to the binding energy, which, in its turn, corresponds to the maximum energy that the asteroids can exchange. In other words, larger separation-binaries can not \corr{provide enough energy exchange between its members in order to allow one of them to became} permanently captured.
\end{enumerate}

%
%

\begin{figure*}
 \includegraphics[width=0.7\textwidth,viewport=0 10 380 230, clip]{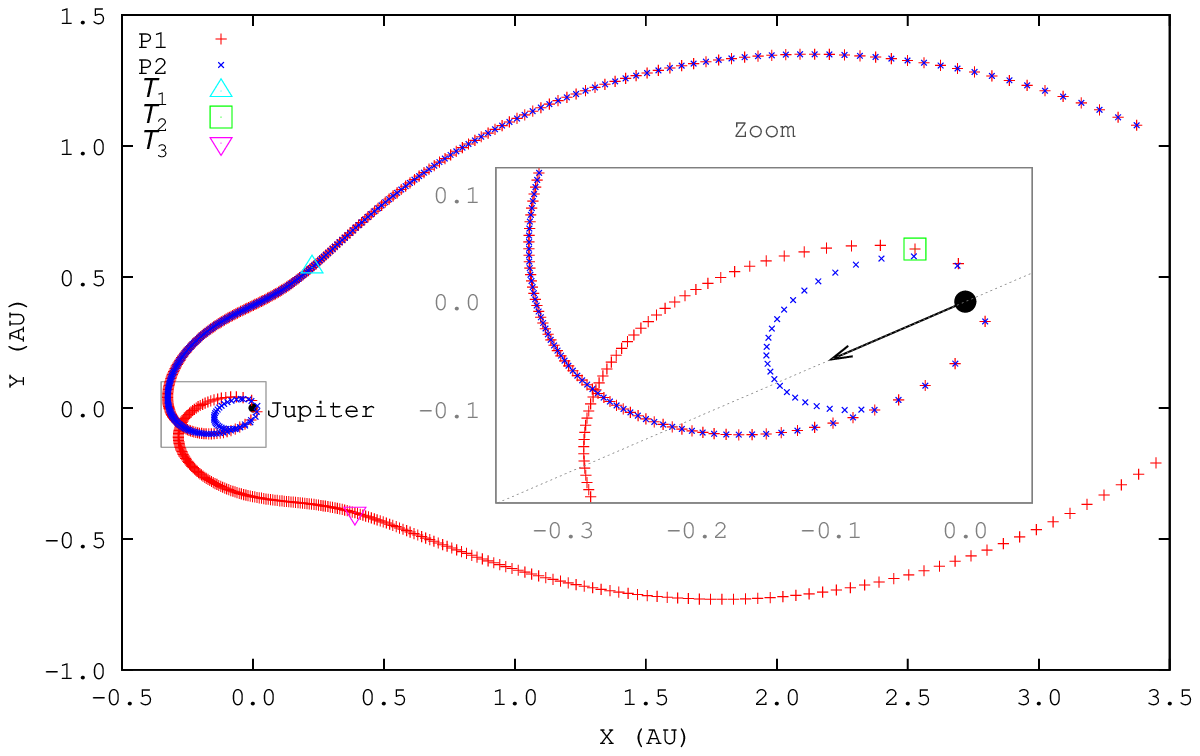}(a)\\
 \includegraphics[width=0.7\textwidth,viewport=0 30 380 220, clip]{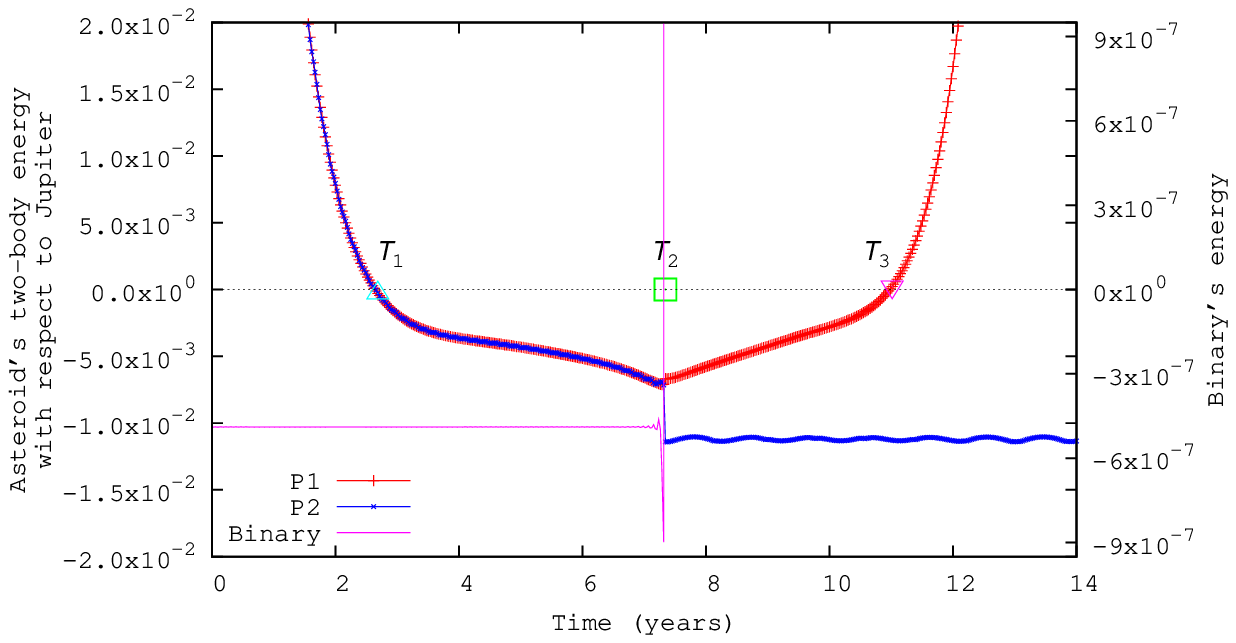}(b)\\
 \includegraphics[width=0.7\textwidth,viewport=0 10 450 270, clip]{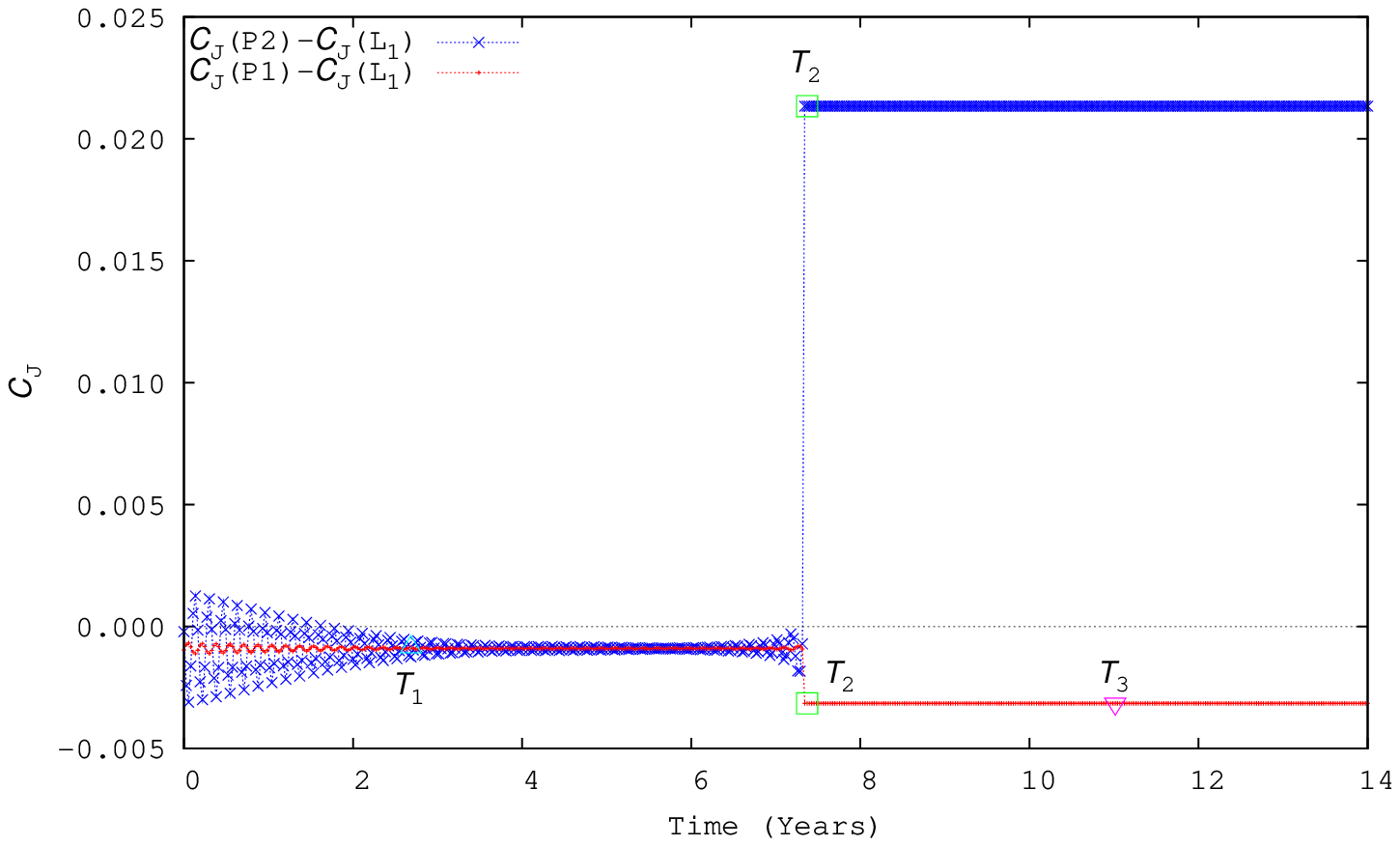}(c)\\
 \caption{A first example of capture process. In plot (a), red \textquoteleft{+}\textquoteright\ and blue \textquoteleft{x}\textquoteright\ show the trajectories of P1 and P2 in Jupiter's (black circle) \corr[20a]{planetocentric non-rotating} frame of reference, respectively. \gram{The output frequency is 10 days}. \corr{Dashed grey line and the black arrow, in the zoom window, indicate Sun direction at the disruption instant (\Tii)}. Tracking binary-asteroid trajectory, one sees that it becomes temporarily captured (light blue triangle) by Jupiter, disrupts (green square inside zoom box) and finally has its minor member permanently captured by Jupiter while its major member escapes from Jupiter's vicinity (pink down triangle). Plots (b) and (c) show the time evolution of energy and $C_J$ for the trajectory of plot (a), respectively. In plot (b), red \textquoteleft{+}\textquoteright\ and blue \textquoteleft{x}\textquoteright\ are the two-body energies of P1 and P2 with respect to Jupiter, respectively, and the pink filled line is the binary's two-body energy, i.e., P2's two-body energy with respect to P1. In plot (c), red \textquoteleft{+}\textquoteright\ is the difference of $C_j$ values calculated for asteroid P1 and $L_1$ Lagrangian point, and similarly , blue \textquoteleft{x}\textquoteright\ difference for P2 and $L_1$.}
 \label{fig:TrajSample}
\end{figure*}

Plot (b) of \refig{TrajSample} illustrates this energy exchange process for the trajectory shown in plot (a). \corr{Plot (c) shows the time evolution of the Jacobi constant value for each individual asteroid. It illustrates that shortly after instant \Tii\ the interaction between P1 and P2 becomes negligible. Note also that, the Jacobi constant value of P2 higher than Jacobi constant value of Lagrangian point $L_1$. Therefore, shows that P2 will never escape Jupiter's vicinity}   

\corr[18c]{Nevertheless, we could expect this increasing fraction of captured low-semimajor axis binaries to have a maximum where it must roll over since very tighter binaries should reach a limit in which the binary-asteroids can be considered as a particle and would never disrupt. Furthermore, we should also expect an increasing fraction of mutual collisions inasmuch binary semi-major axis decreases.}

\subsection{Analysis at the instant of binary disruption (\Tii)}
As defined before, the instant \Tii\ is characterized by the binary-asteroid disruption. So, the graphics in this subsection refer to elements of each asteroid individually with respect to Jupiter.

\begin{figure}
\begin{flushright}
	\includegraphics[width=0.48\textwidth, viewport= 33 15 450 280, clip]{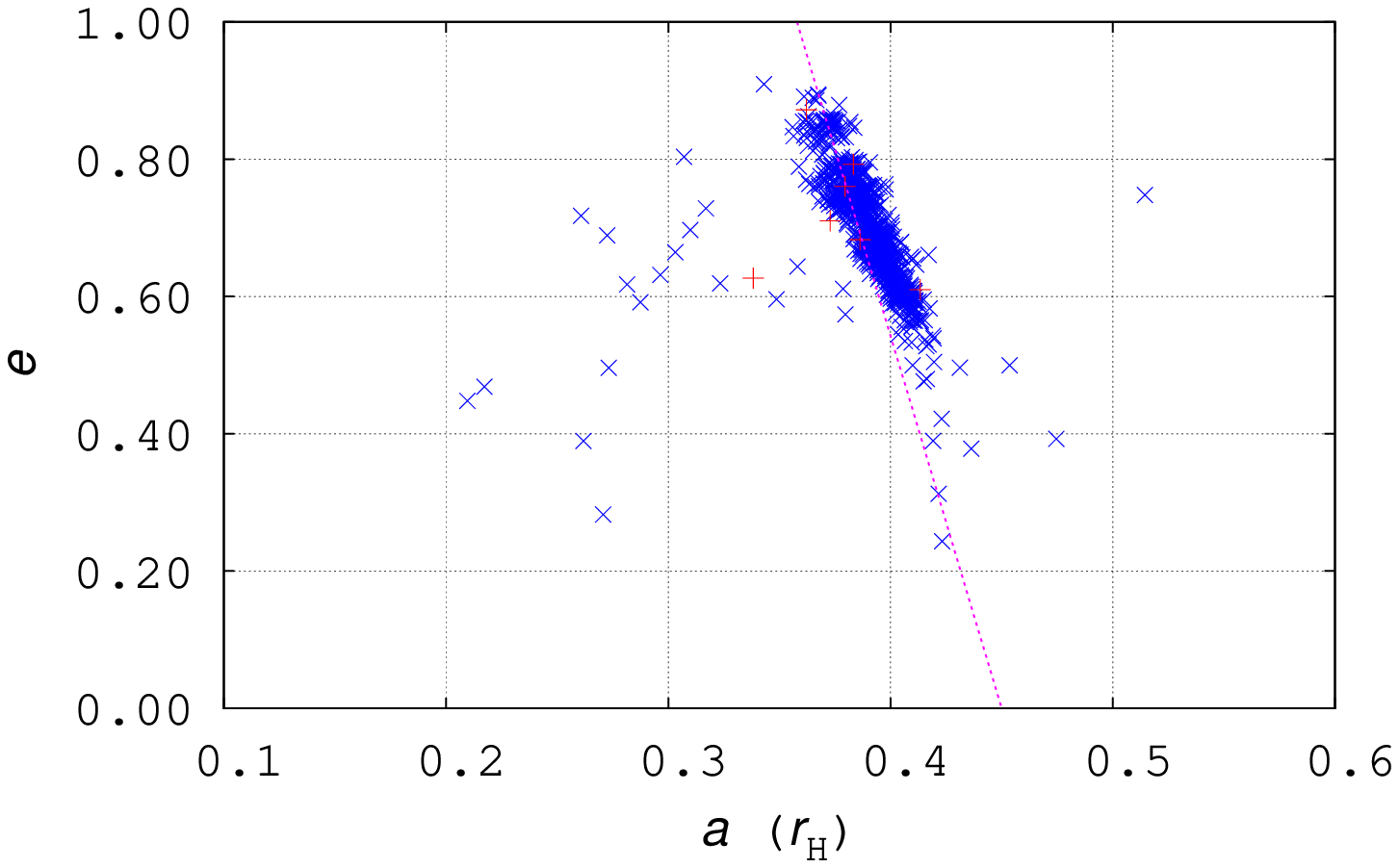}\\
	(a)\\
	\includegraphics[width=0.48\textwidth, viewport= 33 15 450 290, clip]{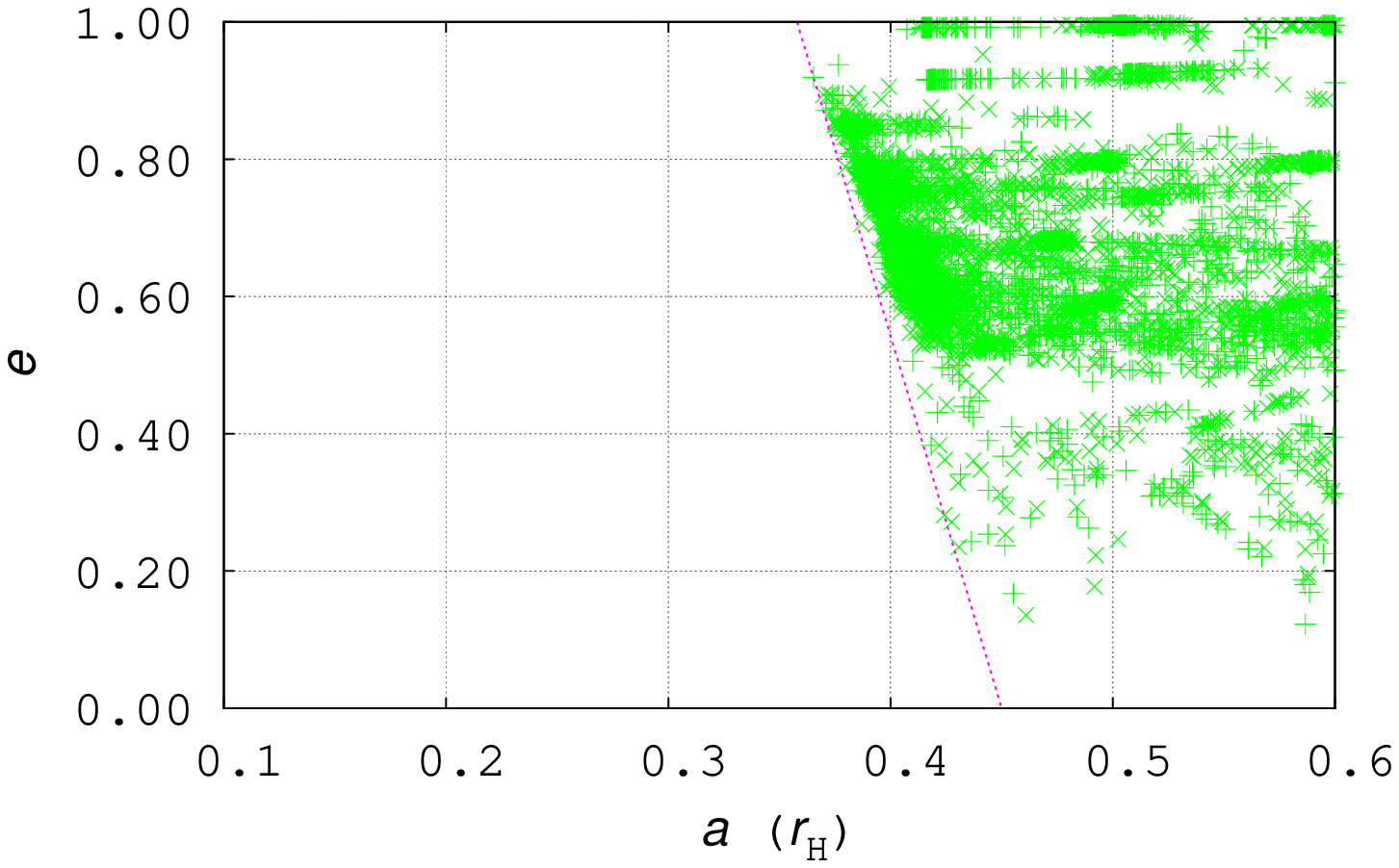}\\
	(b)
\end{flushright}
	\caption{Diagrams of semi-major axis versus eccentricity of the asteroids at the instant \Tii. Plot (a) shows the cases which resulted in permanent captures of either P1 or P2, in red and blue, respectively. Plot (b) shows the cases which resulted in double escapes. The green \textquoteleft{+}\textquoteright\ corresponds to the orbital elements of the asteroids that escape after instant \Tiii. \corr[25]{Pink dotted straight line is the fitted limit semi-major axis given by \refeq{alimit}.}}
	\label{fig:Tiiaxe}
\end {figure}

Plot (a) in the \refig{Tiiaxe} is an $a{\times}e$ diagram, at the instant \Tii, of the asteroids that remained permanently captured by Jupiter. Plot (b) shows the $a{\times}e$ diagram of the last asteroid to escape from Jupiter for double escape cases. By comparing these two diagrams, one finds a region on $a{\times}e$ space in which an asteroid remains permanently captured if its semi-major axis and eccentricity are enclosed within it at instant \Tii. \corr[24]{It must be clear that \refig{Tiiaxe} shows \Tii\ instantaneous diagram of osculating elements which must vary in time due to solar perturbation. Furthermore, we should expect some weak interaction between the pair so far as they recede sufficiently away from each other. By checking the variation of Jacobi constant value, of each individual asteroid, we could estimate how long it takes to this mutual interaction become negligible. By considering a variation of the order of $10^{-8}$ in Jacobi constant value,  we found that for the worst case it took about half orbital period about Jupiter ($\sim$ 170 days) to satisfy the condition.}

 By fitting an expression that bounds this region in the $a{\times}e$ space we found a limit on semi-major axis given in terms of eccentricity, as follow:

\begin{equation}
	a^*(e) = 0.4500(1.0000 - 0.2046e)
	\label{eq:alimit}
\end{equation}

\corr[26a]{The condition $ a < a^*(e)$ at instant \Tii\ can be thought as sufficient but not necessary capture condition. In other words, if the captured object obeys $a < a^*(e)$, then our simulations show that the temporary capture always becomes permanent. However, we also found large number of permanent captures that had a slightly greater semi-major axis in a region where there is a mix of captures and double escapes}

\begin{figure}
	\includegraphics[width=0.48\textwidth,clip]{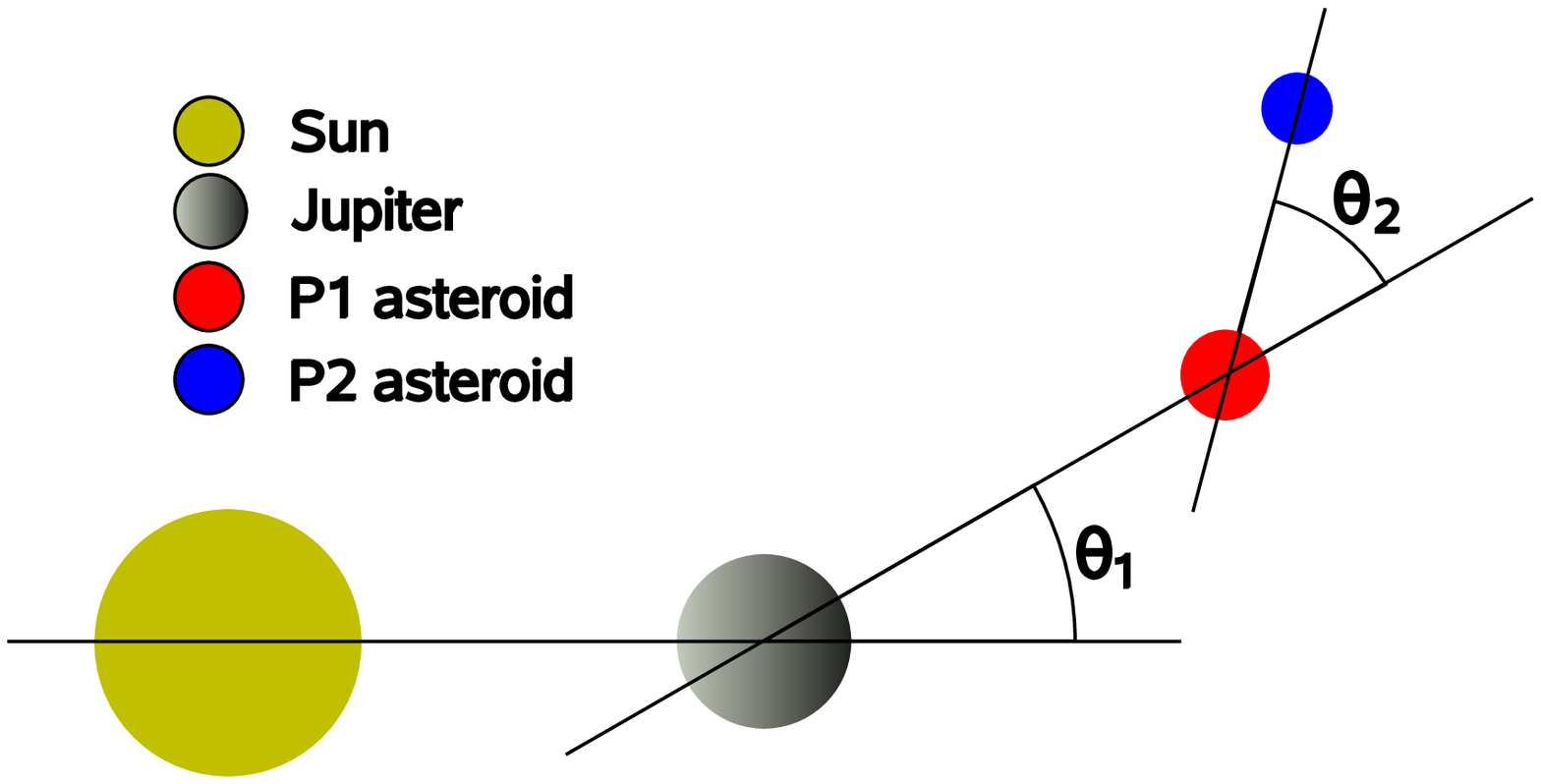}
	\caption{Sketch of angular configuration between the bodies. One can infers a Sun-Jupiter-P1 alignment if \tetai=$0$ or \tetai=$180$\dgr and a Jupiter-P1-P2 alignment when \tetaii=$0$ or \tetaii=$180$\dgr.}
	\label{fig:angularsketch}
\end{figure}

A second graphic at instant \Tii\ is as an angular illustrative histogram. \corr[27]{One can better understand the angular configuration analyzing the phase angles \tetai\ and \tetaii\ shown in \refig{angularsketch}. The angle \tetai\ gives the P1 phase angle with respect to the Sun-Jupiter direction, while, \tetaii\ gives P2 phase angle with respect to the Jupiter-P1 directions. Note that in this system P1 orbits counterclockwise Jupiter and P2 orbits counterclockwise P1.}

\begin{figure}
\begin{flushright}
	\includegraphics[width=0.48\textwidth,clip]{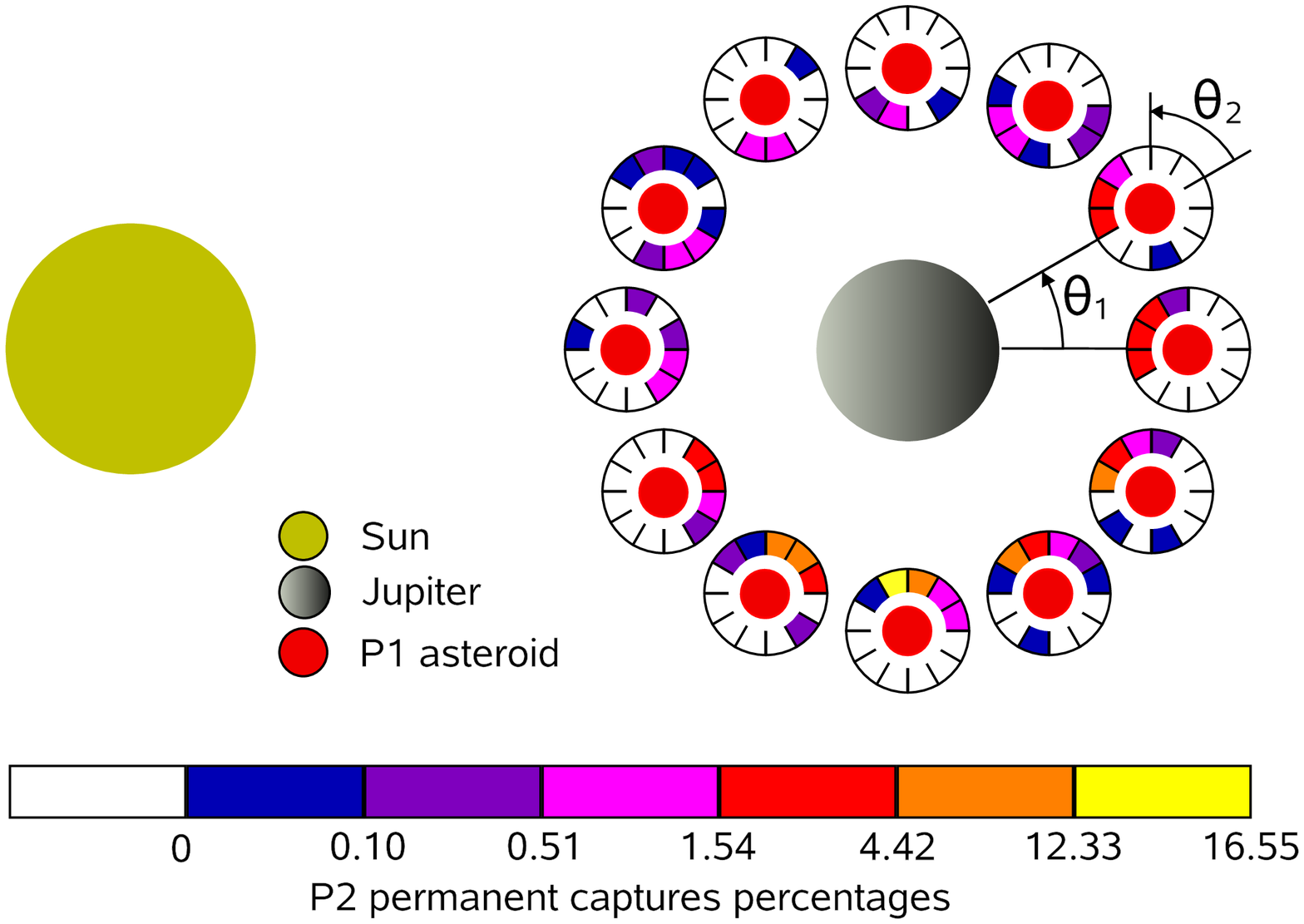}\\
	(a)\\
	\includegraphics[width=0.48\textwidth,clip]{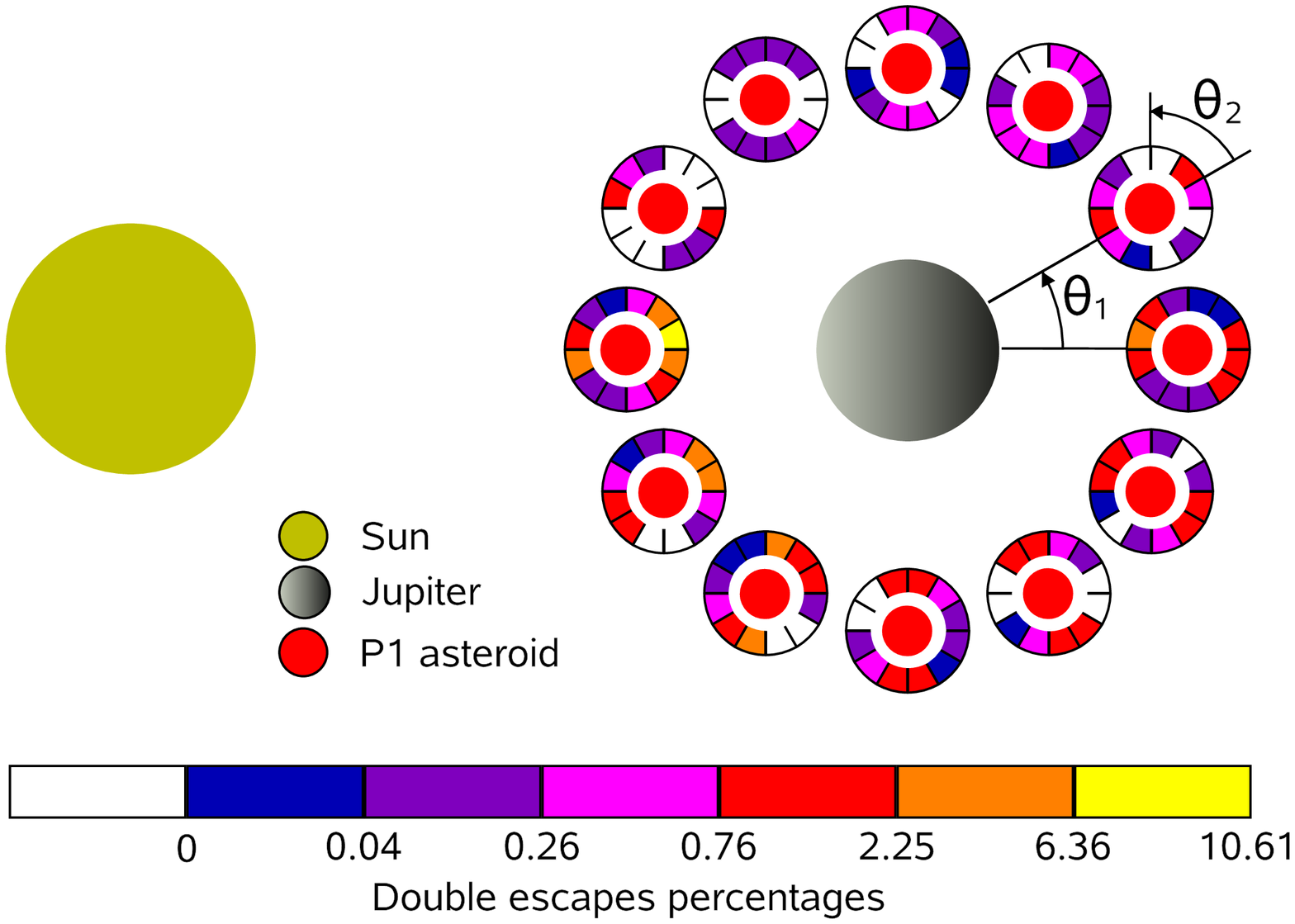}\\
	(b)
\end{flushright}
	\caption{Angular histograms of instant \Tii. The red circles indicate the angular position of P1 with respect to Sun-Jupiter direction given by angle \tetai. The angular sections that surround each red circle indicate the angular position of P2 with respect to Jupiter-P1 direction, given by \tetaii\ \corr[28b]{(see \refig{angularsketch})}. The color of the angular sections in histogram (a) correspond to the percentages of the total number of P2's permanent captures (972; see \reftab{Percentages}), and the color of angular sections in histogram (b) correspond to the cases which resulted in double escapes \corr[28c]{(7446;~\reftab{Percentages})}, given by the respective color scales. White color indicates absolute null number of events. The bodies orbit counterclockwise.}
	\label{fig:angularhistograms}
\end{figure}

Analysing the angular histograms of \refig{angularhistograms} we see that: i) the disruption preferentially occurs when Jupiter, P1 and P2 are approximately aligned, i.e., $\mtetaii\approx 0$ or $\mtetaii\approx180$\dgr. The \corr[28a]{smaller probability for $\mtetai\approx180$\dgr\ in capture histogram (a), plus the higher probability observed for $\mtetai\approx180$\dgr\ in escape histogram (b)} indicate that ii) disruptions which occurs when binary-asteroid is located aligned between Jupiter and Sun most likely result in double escapes. On the other hand, histogram (a) shows that iii) permanent capture of P2 most likely results from binary-asteroids which disrupt at a angular position approximately 90\dgr\ after it cross the Sun-Jupiter line ($\mtetai\approx 270$\dgr). Finally, from the capture histogram (a) one also notes that permanent captures of P2 succeed from disruptions which occurred when P2 were at inferior conjunction with P1, as seen from Jupiter ($\mtetaii\approx 180$\dgr).

\corr[28d]{The preference for captures with the smaller asteroid unbinding when closer to Jupiter ($\mtetaii\approx 180$\dgr) is simply understood as being due to the velocity vector about its center-of-mass motion being opposite to the planetocentric velocity in that geometry. Consequently, asteroid  P2 gets its individual speed, with respect to Jupiter, reduced to a minimum.}

\corr[28e]{The preference for captures when $\mtetai \approx 270$\dgr\ is understood as being due to the solar perturbation, i.e., the Sun's gravity tends to increase the velocity of binary center of mass about Jupiter when $\mtetai \approx 90$\dgr\ while it tends to decrease this velocity when $\mtetai \approx 270$\dgr.} These characteristic angular positions with respect to the Sun-Jupiter line, reinforce the importance of the solar presence on the dynamics of binary-asteroid captures: 

\subsection{Analysis at the escape instant of one asteroid (\Tiii)}
\label{subsec:ResultsTiii}
\begin{figure*}
	\includegraphics[width=0.8\textwidth,viewport=0 10 360 260, clip]{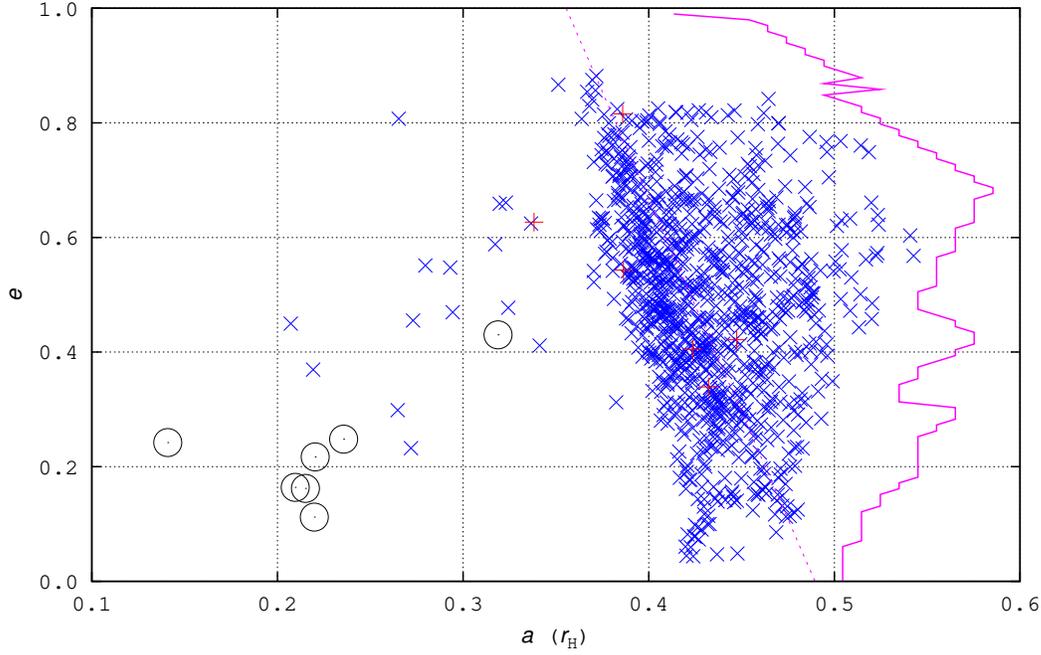}
	\caption{Same as plot (a) in \refig{Tiiaxe} for instant \Tiii, though. Red \textquoteleft{+}\textquoteright\ and blue \textquoteleft{x}\textquoteright\ represent the cases which resulted in permanent captures of either P1 or P2, respectively. Black circumferences represent the orbital elements of Jupiter's real prograde irregular satellites. The pink dotted line is the same as in plot (a) of \refig{Tiiaxe}. The filled pink contour is a stability edge we found numerically as discussed in section \ref{subsec:ResultsTiii}.}
	\label{fig:Tiiiaxe}
\ 
\end{figure*}

The data stored at instant \Tiii\ shows the final configuration of the captured asteroid, given that the captured asteroid will not suffer any interaction with its primordial partner. \refig{Tiiiaxe} is an $a\times e$ diagram of captured asteroid at instant \Tiii. 
The pink filled contour is an extended border of a more general region of stability. In fact, we firstly worried about the points located beyond critical semi-major axis found by \citet{domingosetal06} since they represent the permanently captured asteroids. Nevertheless, by performing a more general capture time analysis we found the extended stability border.

\begin{figure*}
\begin{flushright}
	\includegraphics[width=0.48\textwidth,viewport=17 16 402 340, clip]{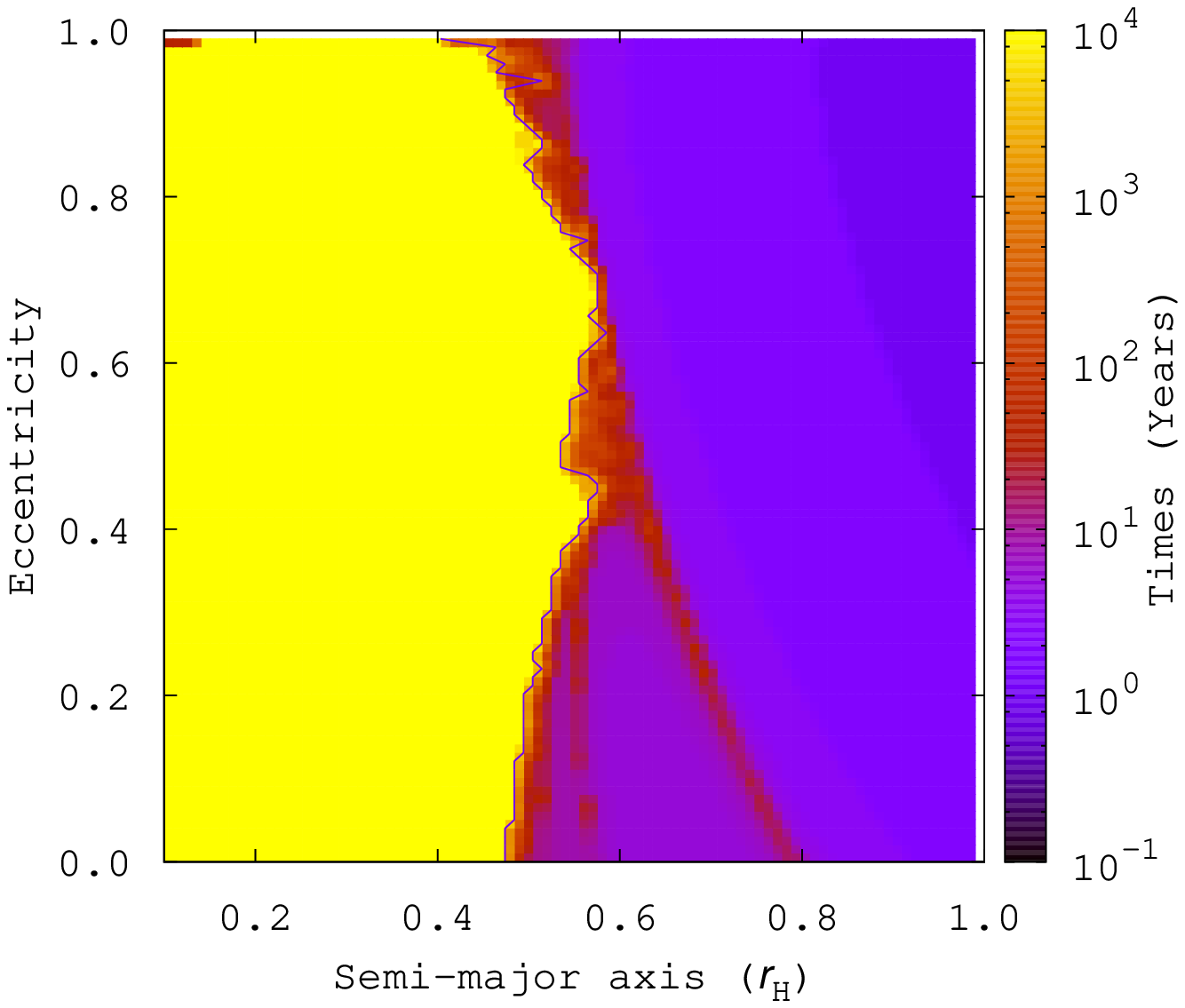}
	\includegraphics[width=0.48\textwidth,viewport=13 16 401 335, clip]{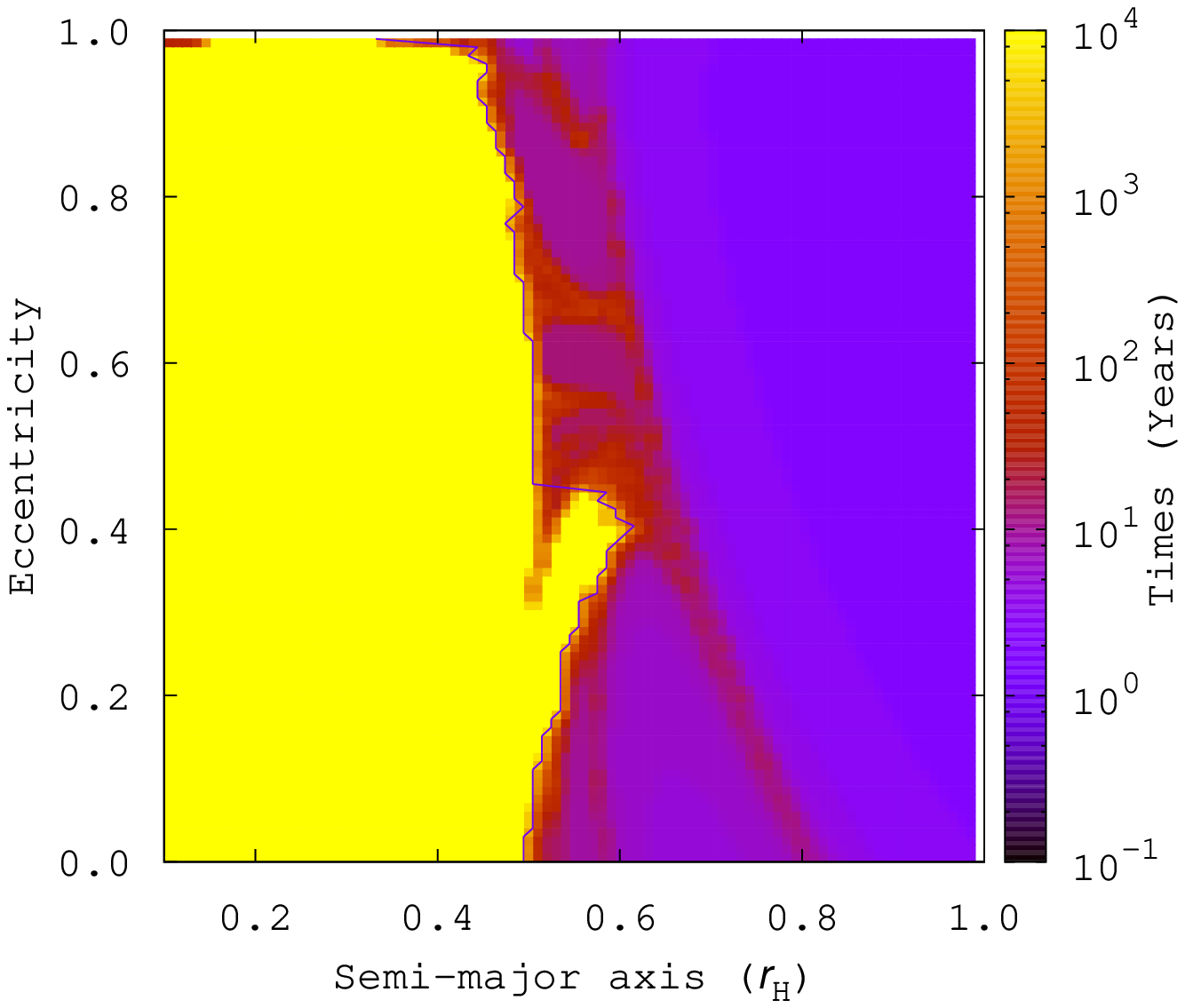}\\
	(a)\hspace{.47\textwidth}(b)
\end{flushright}
	\caption{Same as \refig{Mapa} for (a) $\omega_0=0$ and $\lambda_0=180$\dgr\ and (b) $\omega_0=180$\dgr\ and $\lambda_0=180$\dgr.}
	\label{fig:Mapas}
\end{figure*}

We performed this more general study following the same procedure described at section \ref{subsec:Capturetime}, but considering distinct initial values for longitude of pericenter $\varpi_0$ and true anomaly $f_0$. From the results, shown in the \refig{Mapas}, we found that the pairs of initial values ($\varpi_0=0,f_0=180$\dgr) and ($\varpi_0=180$\dgr$,f_0=180$\dgr) yield capture time maps in which the stability regions are much larger than for initial values ($\varpi_0=0,f_0=0$), shown in \refig{Mapa}. By combining the borders of both maps, in such manner we obtain the larger region of stability, we built the referred more general stability edge shown, as  a pink filled line, in \refig{Tiiiaxe}.

The points in the plot of \refig{Tiiiaxe} shows that the final orbital elements of permanently captured asteroids cover a wide region in the $a\times e$ space. \corr[29]{Most of the captured objects are very far from the planet ($\mofbin{a} \gtrsim 0.4 \mrH$) and will probably be removed due to perturbations not included in our study. The good candidates to survive are those closer to the planet ($\mofbin{a} \lesssim 0.35 \mrH$). Since in this study we considered only the prograde planar case, we are not able to compare our results with the currently known Jupiter's irregular satellites. However, we included them in the plot of \refig{Tiiiaxe} just to have an idea of the orbital shape of captured objects. In order to make a fair comparison it will be needed to make a study in the 3-D space considering the inclinations. This is a study in progress}
%
%

\subsection{Binary captures in numbers}
\begin{table}
 \centering
 \begin{minipage}{0.48\textwidth}
  \caption{Percentages of captures and collisions.}
\label{tab:Percentages}
   \begin{tabular}{@{}lrr@{}}
	\hline
	Description	&Short capture times	&Long capture times\\
	\hline
	Simulations		&3\,888			&4\,860	\\
	Captures		&7			&972	\\
	Captures of P1		&1			&5	\\
	Captures of P2		&6			&966	\\
	Double captures\footnote{P1 and P2 captured}
				&0			&1	\\
	Collisions		&180			&143	\\
 	\hline
\end{tabular}
\end{minipage}
\end{table}

Among 8\,748 simulated trajectories, 4\,860 are from long capture time primary initial conditions and 3\,888 from short capture time ones. \reftab{Percentages} presents the quantities of permanent captures and collisions. Second and third columns show the values with respect to the short and long capture time cases, respectively.
As it is shown in \reftab{Percentages}, though permanent capture probability of the cases derived from short\gram{-}time primary initial conditions are low (0.18\,\%), \gram{the} permanent capture probability of cases derived from long time conditions are \gram{much larger} (20\,\%). The collision probabilities for both long and short times derived from primary initial conditions have the same order and are not negligible.

\subsection{Sample of capture trajectories}

\begin{figure*}
 \includegraphics[width=0.8\textwidth,clip]{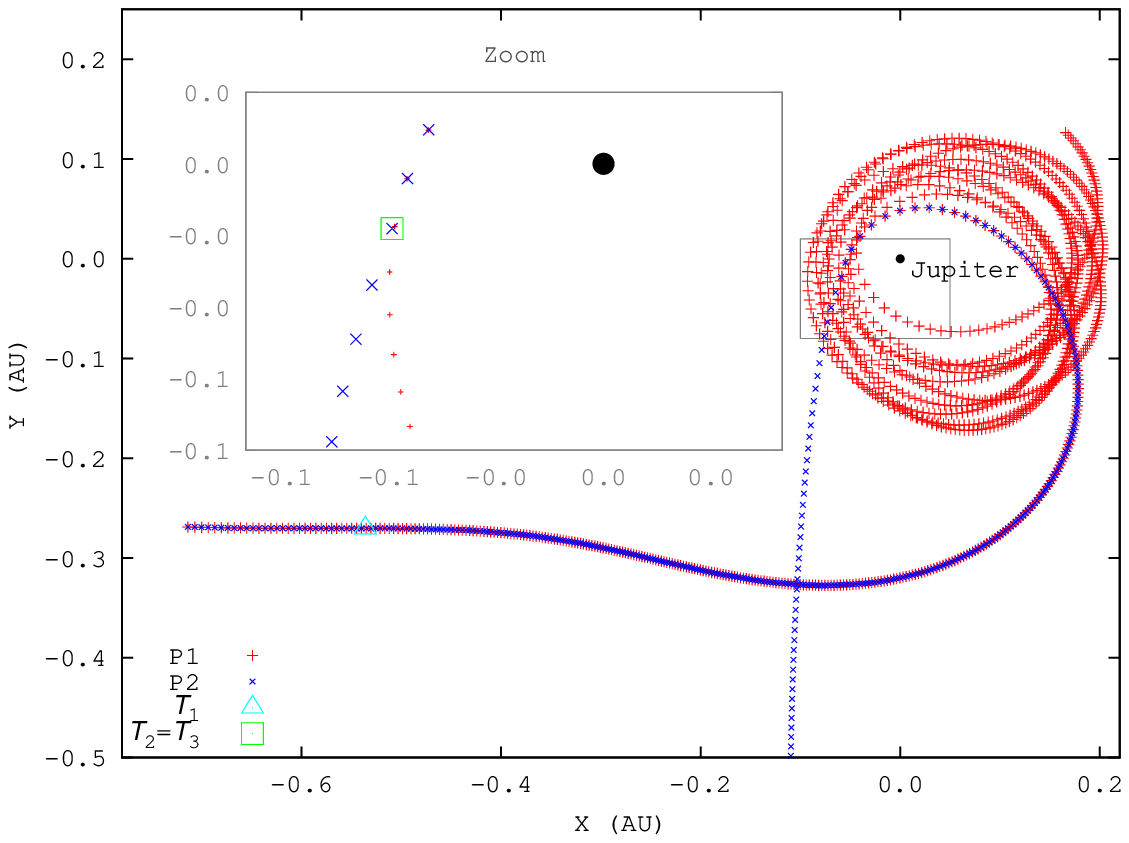}(a)
 \includegraphics[width=0.8\textwidth,clip]{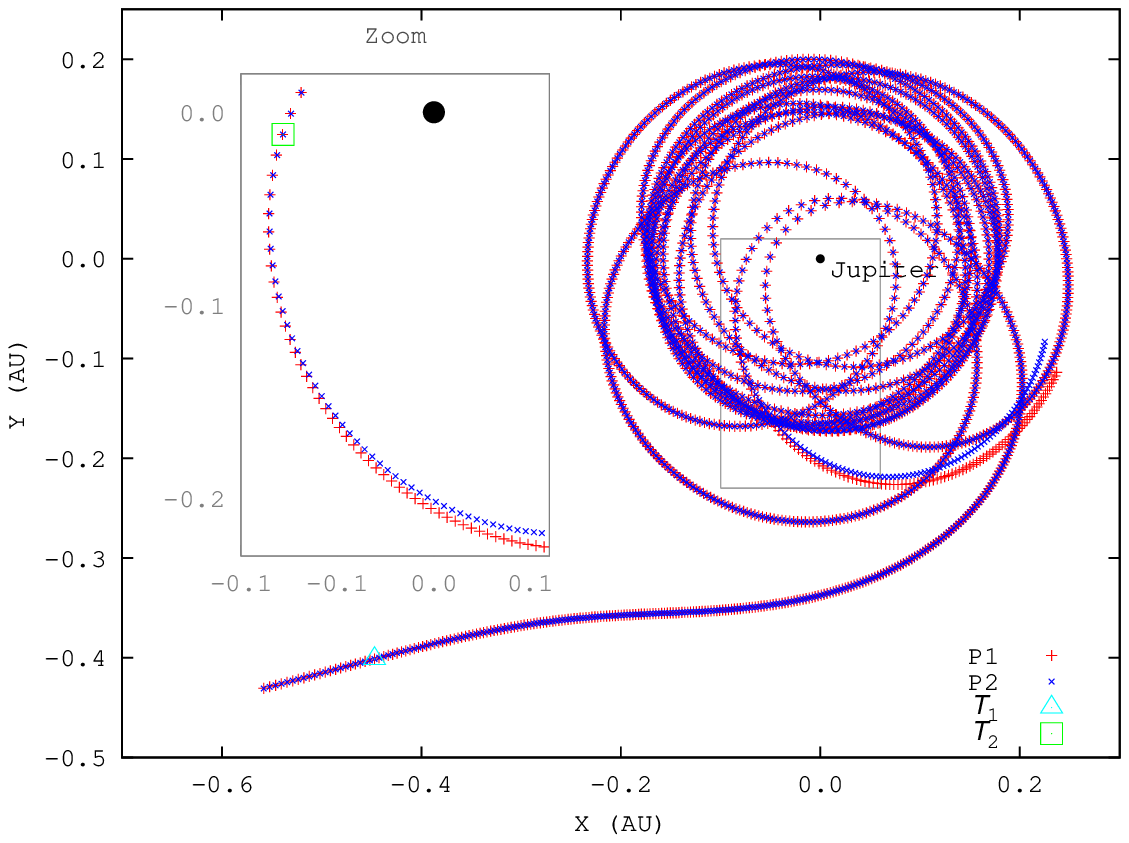}(b)
 \caption{Two examples of capture pathways. Red \textquoteleft{+}\textquoteright\ and blue \textquoteleft{x}\textquoteright\ are the coordinates of P1 and P2, respectively. Light blue triangle and green square denote \Ti and \Tii, respectively. (a) The asteroid P1 remains permanently captured by Jupiter. (b) Both asteroids remain captured by Jupiter even after disruption.}
 \label{fig:TrajSamples}
\end{figure*}

This section presents some examples of capture trajectories \gram{from} three distinct cases. Firstly, \gram{we already showed} in \refig{TrajSample}(a), a typical example where a binary-asteroid approached Jupiter, became captured, disrupted after a while and had its minor member permanently captured by Jupiter while its major member escaped Jupiter's vicinity. \refig{TrajSamples}(a), shows \gram{rare} example where the major asteroid remains captured by Jupiter after disruption of the primordial binary-asteroid. Furthermore, another peculiarity in this example is \Tii=\Tiii, i.e., \gram{the} escape of P2 happens at the instant of binary disruption. Finally, \refig{TrajSamples}(b) presents \gram{our} single case \gram{among $>8000$ performed} where both asteroids remained captured by Jupiter up to the end of integration, even after the disruption of the primordial binary-asteroid. \corr[30]{In both examples of the \refig{TrajSamples} we simulated the trajectories for $10^4$ years, and for both cases the instant \Tii\ happens before $10^2$ years of integration. Nevertheless, by checking the Jacobi constant value for each one of the bodies of the double capture example (\refig{TrajSamples}b), we found that their values are smaller than the $L_1$ Jacobi constant value. Therefore, these bodies may eventually escape from the planet.}

\section{Conclusions}
\label{sec:conclusions}
In this work we have studied the capture dynamics of binary-asteroids by looking for favorable conditions of capture. Our results present new perspectives about the problem of binary-asteroid captures emphasizing the importance of Sun's role in the dynamics. \corr[31]{The Sun is not necessary to produce a binary rupture since Jupiter alone can do it. However, the Sun plays a key role in the disruption process in order to make the capture to become permanent.} The results allow us to comprehend about both binary-asteroid's features as well as intrinsic features of capture process' main stages.

The observed characteristics at the first main stage, \Ti, have revealed that: i) Tighter binary-asteroids are more susceptible to permanent captures than binaries with larger separation. In fact, the permanent capture probability behaves inversely proportional to the binary's semi-major axis. \corr[NOVO]{This results indicate that binary's energy exchange allows the asteroid to become permanently captured. That is, since tighter binaries interactions are more intense, its members can exchange higher amount of energy. In such a way one asteroid can have its energy sufficiently decreased in order to not be able to escape from Jupiter.} \corr[14]{Through this conclusion, we could argue that binary-asteroids with high eccentricities would disrupt more easily, but would not exchange the necessary amount of energy to result in a permanent capture.}\corr[32]{As mentioned, there must exist a lower binary-separation limit below which the binary never disrupts and consequently there is no capture.}

The observed characteristics at the second main stage, \Tii, tell about process' features. It was shown that the angular position of bodies at disruption instant (\Tii) are related with the permanent capture probability. Summarizing: ii) Disruption preferentially occurs when both asteroids are approximately aligned with Jupiter; Nevertheless, iii) disruptions which occurs when P2 is located between Jupiter and P1 result more often in permanent capture; Finally, we found that iv) the permanent capture probability is higher when the binary-asteroid disrupts in an \ eastern quadrature\, i.e., an angular position approximately 90\dgr\ after the binary cross the Sun-Jupiter direction.

The permanent capture probability for an specific set of initial conditions, such derived from long time primary initial conditions, was shown to reach 20\,\%. Therefore, the good candidates are those derived from long capture time primary initial conditions.

As a note, we give here a reference of a similar work submitted to Icarus journal, which also considers the solar perturbation. This work is available on astro-ph \citep{philpottetal_arXiv09}.

\corr[33]{Finally, as the main goal of this paper was to address the favorable conditions which makes the permanent capture plausible, we have chosen a procedure to get initial conditions without taking into account where the incoming objects came from. It means that, in this work we have just tried the model plausibility without taking into account how it could reproduce the currently observed objects. So in order to get a more realistic probability, it must be considered several aspects as inclinations, mass ratios, binary eccentricities as well as to study how realistic are the trajectories of incoming objects.}
\section*{Acknowledgments}
The comments and questions of an anonymous referee helped to significantly improve this paper. The authors gratefully acknowledge CNPq, FAPESP and CAPES, which have funded this work.
\bsp

\label{lastpage}

\begin{thebibliography}{}

\bibitem[\protect\citeauthoryear{Agnor \& Hamilton}{Agnor \&
  Hamilton}{2006}]{agnorhamilton06}
Agnor C.~B.,  Hamilton D.~P.,  2006, Nature, 441, 192

\bibitem[\protect\citeauthoryear{Astakhov, Burbanks, Wiggins \&
  Farrelly}{Astakhov et~al.}{2003}]{astakhovetal03}
Astakhov S.~A.,  Burbanks A.~D.,  Wiggins S.,    Farrelly D.,  2003, Nature,
  423, 264

\bibitem[\protect\citeauthoryear{Benner \& Mckinnon}{Benner \&
  Mckinnon}{1995}]{bennermckinnon95}
Benner L.~A.,  Mckinnon W.~B.,  1995, Icarus, 118, 155

\bibitem[\protect\citeauthoryear{Burns}{Burns}{1986}]{burns86}
Burns J.~A.,  1986, The Evolution of Satellite Orbits, 1 edn.
The University of Arizona Press, Tucson, pp 117--158

\bibitem[\protect\citeauthoryear{Canup \& Ward}{Canup \&
  Ward}{2002}]{canupward02}
Canup R.~M.,  Ward W.~R.,  2002, AJ, 124, 3404

\bibitem[\protect\citeauthoryear{Canup \& Ward}{Canup \&
  Ward}{2006}]{canupward06}
Canup R.~M.,  Ward W.~R.,  2006, Nature, 441, 834

\bibitem[\protect\citeauthoryear{Carusi \& Valsecchi}{Carusi \&
  Valsecchi}{1979}]{carusivalsecchi79}
Carusi A.,  Valsecchi G.,  1979, Numerical Simulations of Close Encounters
  Between Jupiter and Minor Bodies, 2 edn.
The University of Arizona Press, Tucson, pp 391--416

\bibitem[\protect\citeauthoryear{Colombo \& Franklin}{Colombo \&
  Franklin}{1971}]{colombofranklin71}
Colombo G.,  Franklin F.~A.,  1971, Icarus, 15, 186

\bibitem[\protect\citeauthoryear{{\'C}uk \& Burns}{{\'C}uk \&
  Burns}{2004}]{cukburns04}
{\'C}uk M.,  Burns J.~A.,  2004, Icarus, 167, 369

\bibitem[\protect\citeauthoryear{Domingos, Winter \& Yokoyama}{Domingos
  et~al.}{2006}]{domingosetal06}
Domingos R.~C.,  Winter O.~C.,    Yokoyama T.,  2006, MNRAS, 373, 1227

\bibitem[\protect\citeauthoryear{{Emelyanov}}{{Emelyanov}}{2005}]{emelyanov05}
{Emelyanov} N.,  2005, A\&A, 438, L33

\bibitem[\protect\citeauthoryear{Everhart}{Everhart}{1973}]{everhart73}
Everhart E.,  1973, AJ, 78, 316

\bibitem[\protect\citeauthoryear{Everhart}{Everhart}{1985}]{everhart85}
Everhart E.,  1985, in {Carusi} A.,  {Valsecchi} G.~B.,  eds, Dynamics of
  Comets: Their Origin and Evolution, Proceedings of IAU Colloq. 83, held in
  Rome, Italy, June 11-15, 1984. Edited by Andrea Carusi and Giovanni B.
  Valsecchi. Dordrecht: Reidel, Astrophysics and Space Science Library. Volume
  115, 1985,, p.185 {An efficient integrator that uses Gauss-Radau spacings}.
pp 185--+

\bibitem[\protect\citeauthoryear{Funato, Makino, Hut, Kokubo \&
  Kinoshita}{Funato et~al.}{2004}]{funatoetal04}
Funato Y.,  Makino J.,  Hut P.,  Kokubo E.,    Kinoshita D.,  2004, Nature,
  427, 518

\bibitem[\protect\citeauthoryear{{Gladman}, {Kavelaars}, {Holman}, {Petit},
  {Scholl}, {Nicholson} \& {Burns}}{{Gladman} et~al.}{2000}]{gladmanetal00}
{Gladman} B.,  {Kavelaars} J.,  {Holman} M.,  {Petit} J.,  {Scholl} H.,
  {Nicholson} P.,    {Burns} J.~A.,  2000, Icarus, 147, 320

\bibitem[\protect\citeauthoryear{{Gladman}, {Kavelaars}, {Holman}, {Nicholson},
  {Burns}, {Hergenrother}, {Petit}, {Marsden}, {Jacobson}, {Gray} \&
  {Grav}}{{Gladman} et~al.}{2001}]{gladmanetal01}
{Gladman} B.,  {Kavelaars} J.~J.,  {Holman} M.,  {Nicholson} P.~D.,  {Burns}
  J.~A.,  {Hergenrother} C.~W.,  {Petit} J.-M.,  {Marsden} B.~G.,  {Jacobson}
  R.,  {Gray} W.,    {Grav} T.,  2001, Nature, 412, 163

\bibitem[\protect\citeauthoryear{{Gladman}, {Nicholson}, {Burns}, {Kavelaars},
  {Marsden}, {Williams} \& {Offutt}}{{Gladman} et~al.}{1998}]{gladmanetal98}
{Gladman} B.~J.,  {Nicholson} P.~D.,  {Burns} J.~A.,  {Kavelaars} J.,
  {Marsden} B.~G.,  {Williams} G.~V.,    {Offutt} W.~B.,  1998, Nature, 392,
  897

\bibitem[\protect\citeauthoryear{{Gomes}, {Levison}, {Tsiganis} \&
  {Morbidelli}}{{Gomes} et~al.}{2005}]{gomesetal05}
{Gomes} R.,  {Levison} H.~F.,  {Tsiganis} K.,    {Morbidelli} A.,  2005,
  Nature, 435, 466

\bibitem[\protect\citeauthoryear{Hahn \& Malhotra}{Hahn \&
  Malhotra}{2005}]{hahnmalhotra05}
Hahn J.~M.,  Malhotra R.,  2005, AJ, 130, 2392

\bibitem[\protect\citeauthoryear{{Hamilton} \& {Burns}}{{Hamilton} \&
  {Burns}}{1991}]{hamiltonburns91}
{Hamilton} D.~P.,  {Burns} J.~A.,  1991, Icarus, 92, 118

\bibitem[\protect\citeauthoryear{{Hamilton} \& {Burns}}{{Hamilton} \&
  {Burns}}{1992}]{hamiltonburns92}
{Hamilton} D.~P.,  {Burns} J.~A.,  1992, Icarus, 96, 43

\bibitem[\protect\citeauthoryear{Heppenheimer \& Porco}{Heppenheimer \&
  Porco}{1977}]{heppenheimerporco77}
Heppenheimer T.~A.,  Porco C.,  1977, Icarus, 30, 385

\bibitem[\protect\citeauthoryear{{Holman}, {Kavelaars}, {Grav}, {Gladman},
  {Fraser}, {Milisavljevic}, {Nicholson}, {Burns}, {Carruba}, {Petit},
  {Rousselot}, {Mousis}, {Marsden} \& {Jacobson}}{{Holman}
  et~al.}{2004}]{holmanetal04}
{Holman} M.~J.,  {Kavelaars} J.~J.,  {Grav} T.,  {Gladman} B.~J.,  {Fraser}
  W.~C.,  {Milisavljevic} D.,  {Nicholson} P.~D.,  {Burns} J.~A.,  {Carruba}
  V.,  {Petit} J.,  {Rousselot} P.,  {Mousis} O.,  {Marsden} B.~G.,
  {Jacobson} R.~A.,  2004, Nature, 430, 865

\bibitem[\protect\citeauthoryear{Jewitt \& Haghighipour}{Jewitt \&
  Haghighipour}{2007}]{jewitthaghighipour07}
Jewitt D.~C.,  Haghighipour N.,  2007, Ann. Rev. A\&A, 45, 261

\bibitem[\protect\citeauthoryear{{Kavelaars}, {Holman}, {Grav},
  {Milisavljevic}, {Fraser}, {Gladman}, {Petit}, {Rousselot}, {Mousis} \&
  {Nicholson}}{{Kavelaars} et~al.}{2004}]{kavelaarsetal04}
{Kavelaars} J.~J.,  {Holman} M.~J.,  {Grav} T.,  {Milisavljevic} D.,  {Fraser}
  W.,  {Gladman} B.~J.,  {Petit} J.,  {Rousselot} P.,  {Mousis} O.,
  {Nicholson} P.~D.,  2004, Icarus, 169, 474

\bibitem[\protect\citeauthoryear{Kuiper}{Kuiper}{1956}]{kuiper56}
Kuiper G.,  1956, Vistas in Astronomy, 2, 1631

\bibitem[\protect\citeauthoryear{Lunine \& Stevenson}{Lunine \&
  Stevenson}{1982}]{luninestevenson82}
Lunine J.~I.,  Stevenson D.~J.,  1982, Icarus, 52, 14

\bibitem[\protect\citeauthoryear{Mosqueira \& Estrada}{Mosqueira \&
  Estrada}{2003}]{mosqueiraestrada03}
Mosqueira I.,  Estrada P.~R.,  2003, Icarus, 163, 198

\bibitem[\protect\citeauthoryear{Murray \& Dermott}{Murray \&
  Dermott}{1999}]{murray99}
Murray C.~D.,  Dermott S.~F.,  1999, Solar System Dynamics, 1 edn.
Cambridge University Press

\bibitem[\protect\citeauthoryear{Nesvorn{\'y}, Alvarellos, Dones \&
  Levison}{Nesvorn{\'y} et~al.}{2003}]{nesvornyetal03}
Nesvorn{\'y} D.,  Alvarellos J.~L.,  Dones L.,    Levison H.~F.,  2003, AJ,
  126, 398

\bibitem[\protect\citeauthoryear{Noll}{Noll}{2006}]{noll06}
Noll K.~S.,  2006, in Lazzaro D.,  Ferraz-Mello S.,   Fern{\'a}ndez J.~A.,
  eds, Proc. of IAU: Symp. No. 229 Solar system binaries.
Cambridge University Press, pp 301--318

\bibitem[\protect\citeauthoryear{Oliveira, Winter, {Vieira Neto} \& {de
  Felipe}}{Oliveira et~al.}{2007}]{oliveiraetal07}
Oliveira D.~S.,  Winter O.~C.,  {Vieira Neto} E.,    {de Felipe} G.,  2007,
  EM\&P, 100, 233

\bibitem[\protect\citeauthoryear{Peale}{Peale}{1999}]{peale99}
Peale S.~J.,  1999, Ann. Rev. A\&A, 37, 533

\bibitem[\protect\citeauthoryear{Philpott, Hamilton \& Agnor}{Philpott
  et~al.}{2009}]{philpottetal_arXiv09}
Philpott C.,  Hamilton D.~P.,    Agnor C.~B., , 2009, Three-Body Capture of
  Irregular Satellites: Application to Jupiter, arXiv:0911.1369v1

\bibitem[\protect\citeauthoryear{Pollack, Burns \& Tauber}{Pollack
  et~al.}{1979}]{pollacketal79}
Pollack J.~B.,  Burns J.~A.,    Tauber M.~E.,  1979, Icarus, 37, 587

\bibitem[\protect\citeauthoryear{Pollack, Hubickyj, Bodenheimer, Lissauer,
  Podolak \& Greenzweig}{Pollack et~al.}{1996}]{pollacketal96}
Pollack J.~B.,  Hubickyj O.,  Bodenheimer P.,  Lissauer J.~J.,  Podolak M.,
  Greenzweig Y.,  1996, Icarus, 124, 62

\bibitem[\protect\citeauthoryear{{Sheppard}, {Jewitt} \& {Kleyna}}{{Sheppard}
  et~al.}{2005}]{sheppardetal05}
{Sheppard} S.~S.,  {Jewitt} D.,    {Kleyna} J.,  2005, AJ, 129, 518

\bibitem[\protect\citeauthoryear{{Sheppard}, {Jewitt} \& {Kleyna}}{{Sheppard}
  et~al.}{2006}]{sheppardetal06}
{Sheppard} S.~S.,  {Jewitt} D.,    {Kleyna} J.,  2006, AJ, 132, 171

\bibitem[\protect\citeauthoryear{Sheppard \& Jewitt}{Sheppard \&
  Jewitt}{2003}]{sheppardjewitt03}
Sheppard S.~S.,  Jewitt D.~C.,  2003, Nature, 423, 261

\bibitem[\protect\citeauthoryear{Tisserand}{Tisserand}{1896}]{tisserand1896}
Tisserand F.~F.,  1896, Trait{\'e} de M{\'e}canique C{\'e}leste IV, 1 edn.
Gauthier-Villars

\bibitem[\protect\citeauthoryear{Tsiganis, Gomes, Morbidelli \&
  Levison}{Tsiganis et~al.}{2005}]{tsiganisetal05}
Tsiganis K.,  Gomes R.,  Morbidelli A.,    Levison H.~F.,  2005, Nature, 435,
  459

\bibitem[\protect\citeauthoryear{Tsui}{Tsui}{1999}]{tsui99}
Tsui K.,  1999, Planet.~Space~Sci., 47, 917

\bibitem[\protect\citeauthoryear{Tsui}{Tsui}{2000}]{tsui00}
Tsui K.,  2000, Icarus, 148, 139

\bibitem[\protect\citeauthoryear{{Vieira Neto} \& Winter}{{Vieira Neto} \&
  Winter}{2001}]{vieiranetowinter01}
{Vieira Neto} E.,  Winter O.~C.,  2001, AJ, 122, 440

\bibitem[\protect\citeauthoryear{{Vieira Neto}, Winter \& Yokoyama}{{Vieira
  Neto} et~al.}{2004}]{vieiranetoetal04}
{Vieira Neto} E.,  Winter O.~C.,    Yokoyama T.,  2004, A\&A, 414, 727

\bibitem[\protect\citeauthoryear{Vokrouhlick{\'y}, Nesvorn{\'y} \&
  Levison}{Vokrouhlick{\'y} et~al.}{2008}]{vokrouhlickyetal08}
Vokrouhlick{\'y} D.,  Nesvorn{\'y} D.,    Levison H.~F.,  2008, AJ, 136, 1463

\bibitem[\protect\citeauthoryear{Winter \& {Vieira Neto}}{Winter \& {Vieira
  Neto}}{2001}]{wintervieiraneto01}
Winter O.~C.,  {Vieira Neto} E.,  2001, A\&A, 377, 1119

\end{thebibliography}
\end{document}